\documentclass{aa}
\usepackage{graphicx}
\usepackage[varg]{txfonts}
\usepackage{natbib}
\bibpunct{(}{)}{;}{a}{}{,} 

\usepackage{color}
\usepackage[colorlinks=true,linkcolor=black,citecolor=blue,urlcolor=blue]{hyperref}

\begin{document}
\title{Stellar energetic particle ionization in protoplanetary disks around T~Tauri stars}

\subtitle{}

\author{Ch. Rab, \inst{1}
\and M. Güdel\inst{1}
\and M. Padovani\inst{2}
\and I.~Kamp\inst{3}
\and W.-F.~Thi\inst{4}
\and P.~Woitke\inst{5}
\and G.~Aresu\inst{6}
}

\titlerunning{Stellar energetic particle ionization.}
\authorrunning{Ch. Rab et al.}

\institute{University of Vienna, Dept. of Astrophysics, T\"urkenschanzstr. 17, 1180 Wien, Austria \\
\email{christian.rab@univie.ac.at}
\and INAF-Ossevatorio Astrofisico di Arcetri, Largo E. Fermi, 5 - 50125 Firenze, Italy
\and Kapteyn Astronomical Institute, University of Groningen, P.O. Box 800, 9700 AV Groningen, The Netherlands
\and Max-Planck-Institut für extraterrestrische Physik, Giessenbachstrasse 1, 85748 Garching, Germany
\and SUPA, School of Physics \& Astronomy, University of St. Andrews, North Haugh, St. Andrews KY16 9SS, UK
\and INAF, Osservatorio Astronomico di Cagliari, via della Scienza 5, 09047 Selargius, Italy
}

\date{Received 14 December 2016 / Accepted 24 February 2017}

\abstract
{Anomalies in the abundance measurements of short lived radionuclides in meteorites indicate that the protosolar nebulae was irradiated by a high amount of energetic particles ($\mathrm{E\gtrsim10\,MeV}$). The particle flux of the contemporary Sun cannot explain these anomalies. However, similar to T~Tauri stars the young Sun was more active and probably produced enough high energy particles to explain those anomalies.}   
{We want to study the interaction of stellar energetic particles with the gas component of the disk (i.e. ionization of molecular hydrogen) and identify possible observational tracers of this interaction.}
{We use a 2D radiation thermo-chemical protoplanetary disk code to model a disk representative for T~Tauri stars. We use a particle energy distribution derived from solar flare observations and an enhanced stellar particle flux proposed for T~Tauri stars. For this particle spectrum we calculate the stellar particle ionization rate throughout the disk with an accurate particle transport model. We study the impact of stellar particles for models with varying X-ray and cosmic-ray ionization rates.}
{We find that stellar particle ionization has a significant impact on the abundances of the common disk ionization tracers HCO$^+$ and N$_2$H$^+$, especially in models with low cosmic-ray ionization rates (e.g. $10^{-19}\,\mathrm{s^{-1}}$ for molecular hydrogen). In contrast to cosmic rays and X-rays, stellar particles cannot reach the midplane of the disk. Therefore molecular ions residing in the disk surface layers are more affected by stellar particle ionization than molecular ions tracing the cold layers/midplane of the disk.}
{Spatially resolved observations of molecular ions tracing different vertical layers of the disk allow to disentangle the contribution of stellar particle ionization from other competing ionization sources. Modeling such observations with a model like the one presented here allows to constrain the stellar particle flux  in disks around T~Tauri stars.}
 
\keywords{Stars: formation - Stars: circumstellar matter - Stars: activity - Radiative transfer - Astrochemistry - Methods: numerical}
\maketitle
%
\section{Introduction}
Our Sun acts as a particle accelerator and produces energetic particles with energies $\ge 10~\mathrm{MeV}$ \citep[e.g.][]{Mewaldt2007c}. Such particles are also called "solar cosmic rays" as their energies are comparable to Galactic cosmic rays. They are accelerated in highly violent events like flares and/or close to the solar surface due to shocks produced by coronal mass ejections \citep{Reames2015b}. Therefore the energetic particle flux is strongly correlated with the activity of the Sun \citep[e.g.][]{Mewaldt2005,Reedy2012d}.

From X-ray observations of T~Tauri stars we know that their X-ray luminosities can be up to $10^4$ times higher than the X-ray luminosity of the contemporary Sun \citep[e.g.][]{Feigelson1999,Gudel2007d}. Such high X-ray luminosities are rather a result of enhanced flare activity of young stars than coronal effects \citep[e.g][]{Feigelson2002a}. Enhanced activity of T~Tauri stars implies an increase of their stellar energetic particle (SP) flux. From simple scaling with the X-ray luminosity and considering that young stars produce more powerful flares, \citet{Feigelson2002a} derived a typical SP flux for T~Tauri stars $\approx10^5$ times higher than for the contemporary Sun. Under these assumptions T~Tauri stars might show on average a continuous proton flux of $f_\mathrm{p}(E_\mathrm{p}\ge 10~\mathrm{MeV})\approx10^7\,\mathrm{protons\,cm^{-2}\,s^{-1}}$ at a distance of 1~au from the star.

Such a scenario is also likely for the young Sun. Measurements of decay products of short-lived radionuclides (SLR) like \element[][10]Be or \element[][26]{Al} in meteorites indicate an over-abundance of SLRs in the early phases of our Solar System \citep[e.g.][]{Meyer2000}. One likely explanation for these abundance anomalies are spallation reactions of SPs with the dust in the protosolar nebula \citep[e.g.][]{Lee1998mq,McKeegan2000b,Gounelle2001d,Gounelle2006b}. However, this would require a strongly enhanced SP flux of the young Sun by a factor $\gtrsim3\times10^5$ compared to the contemporary Sun \citep{McKeegan2000b}, consistent with the estimated SP flux of T~Tauri stars derived by \citet{Feigelson2002a}.

More recently \citet{Ceccarelli2014d} reported a first indirect measurement of SPs in the protostar \object{OMC-2 FIR 4}. They observed a low HCO$^+$/N$_2$H$^+$ abundance ratio of $3-4$ that requires high H$_2$ ionization rates $>10^{-14}\,\mathrm{s^{-1}}$ throughout the protostellar envelope. They explain this high ionization rate by the presence of SPs. From this ionization rate they derived a particle flux of $f_\mathrm{p}(E_\mathrm{p}\ge 10~\mathrm{MeV})\ge 3-9\times10^{11}\,\mathrm{protons\,cm^{-2}\,s^{-1}}$ at 1~au distance from the star. Such a high flux would be more than sufficient to explain the over-abundance of SLRs in the solar nebula.

A further indication of SPs in young stars is the anti-correlation of X-ray fluxes with the crystalline mass fraction of the circumstellar dust found by \citet{Glauser2009}. They argue that SPs are responsible for the amorphization of dust particles and that the correlation can be explained if the SP flux scales with the stellar X-ray luminosity. \citet{Trappitsch2015} tested such a scenario by using detailed models of SP transport for a protoplanetary disk (i.e. SP flux as a function of height of the disk). They also considered the vertical "mixing" of dust particles. According to their models SP irradiation of the disk cannot explain the total SLR abundances in the solar nebula but might play a role for dust amorphization.

Like Galactic cosmic rays SPs not only interact with the solid component but also with the gas component of the disk. However, little is known about the impact of SPs on the chemical structure of disks. \citet{Turner2009bb} investigated the relevance of SPs on the size of dead-zones in disks. Assuming similar enhancement factors as mentioned above they find that SPs can decrease the size of the dead zone depending on the disk model and other ionization sources. 

We present a first approach to study the impact of SP ionization on the chemical structure of the disk. We assume a typical SP flux as proposed  for T~Tauri stars to study the impact of SP ionization on the common disk ionization tracers HCO$^+$ and N$_2$H$^+$. We use the radiation thermo-chemical disk model P{\tiny RO}D{\tiny I}M{\tiny O} (PROtoplanetary DIsk MOdel, \citealt{Woitke2009a,Kamp2010,Thi2011,Woitke2016}) to model the thermal and chemical structure of the disk. We argue that spatially resolved radial intensity profiles of molecular ion emission of the disk allow to constrain the SP flux of T~Tauri stars.

In Sect.~\ref{sec:method} we describe our method to derive the ionization rate due to SPs and the disk model. Our results are presented in Sect.~\ref{sec:results} where we show the impact of SPs on the common disk ionization tracers HCO$^+$ and N$_2$H$^+$. In Sect.~\ref{sec:discussion} we discuss possibilities to constrain the SP flux via observations of molecular ion emission and future prospects for modeling of SP ionization in protoplanetary disks. We present a summary and our main conclusions in Sect.~\ref{sec:conclusions}.

\section{Method}
To investigate the impact of SPs on the disk chemical structure we first need to determine the SP flux and the particle energy distribution. With these particle spectra we can calculate the ionization rate throughout the disk. We apply this to a disk structure representative for disks around T~Tauri stars. With the radiation thermo-chemical disk code P{\tiny RO}D{\tiny I}M{\tiny O} we calculate the chemical abundances. We do this for a series of models where we also consider other important high energy ionization sources like Galactic cosmic rays (CR) and X-rays.
\label{sec:method}
\subsection{Stellar energetic particle spectra}
As the actual particle spectra and fluxes of young stars are unknown we derive the  spectra from the knowledge available from our Sun. The origin of solar energetic particles are most likely flares and/or shock waves driven by coronal-mass ejections (CME) \citep{Reames2013,Reames2015b}. Flares act like point sources on the solar surface whereas the shock waves can fill half of the heliosphere at around 2~Solar radii \citep{Reames2015b}. Particle fluxes are not continuous but rather produced in events lasting from several hours to days \citep{Feigelson2002a,Mewaldt2005}. 

Based on observed X-ray luminosities of solar analogs in the Orion nebula \citet{Feigelson2002a} estimated that SP fluxes in young stars are likely $\approx10^5$ times higher than in the contemporary Sun (see also \citealt{Glassgold2005a}). As T~Tauri stars are very active \citet{Feigelson2002a} argue that it is likely that X-ray flares with luminosities below the detection limit occur several times a day (the same argument holds for CMEs). In that case the X-ray flares and consequently also SP events overlap, resulting in an enhanced continuous SP flux.

Based on these arguments we assume here a continuous and enhanced SP flux for young T~Tauri stars. This approximation is consistent with the assumption of powerful and overlapping flare and CME events of T~Tauri stars.

\begin{figure}
\resizebox{\hsize}{!}{\includegraphics{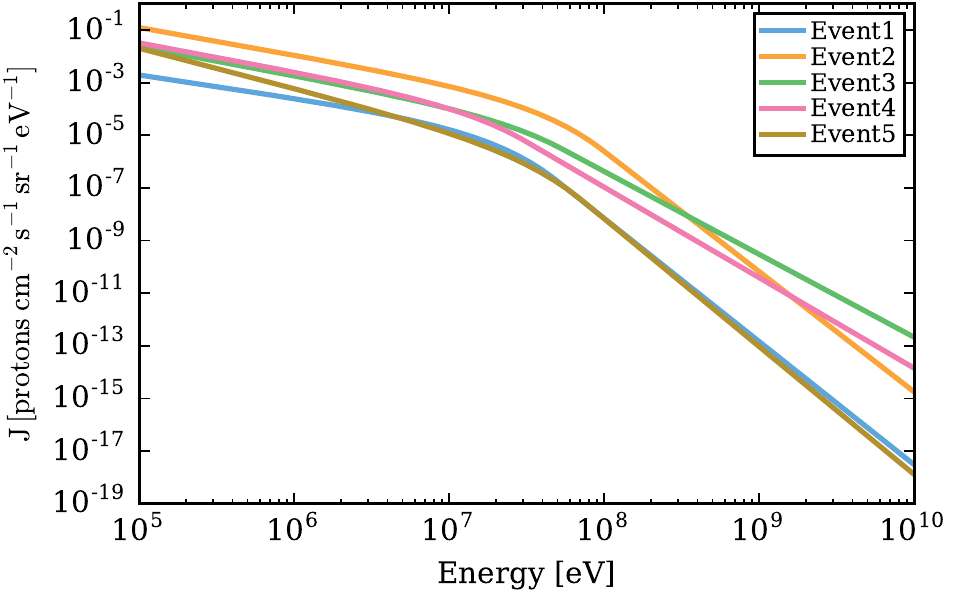}}
\caption{Stellar particle (proton) spectra for five different solar particle events. Shown are fits to the measurements presented in \citet{Mewaldt2005}.}
\label{fig:spectrasun}
\end{figure}
In \citet{Mewaldt2005} measurements of 5 different solar particle events are reported. We use their fitting formulae (see their Eq.~(2) and Table~5) and derive SP spectra (protons in this case) averaged over the duration of the observed events. These measurements are for particles with energies up to several 100~MeV. We extrapolate their results up to energies typical for Galactic cosmic rays of $\approx10\,\mathrm{GeV}$. This is consistent with the maximum energy $E_\mathrm{max}\approx30\,\mathrm{GeV}$ derived by \citet{Padovani2015f,Padovani2016b} for particles accelerated on protostellar surfaces. As seen from Fig.~\ref{fig:spectrasun} the flux levels for the different events can vary by up to 2 orders of magnitudes and there is also some variation in the shape of the spectra.

\begin{figure}
\resizebox{\hsize}{!}{\includegraphics{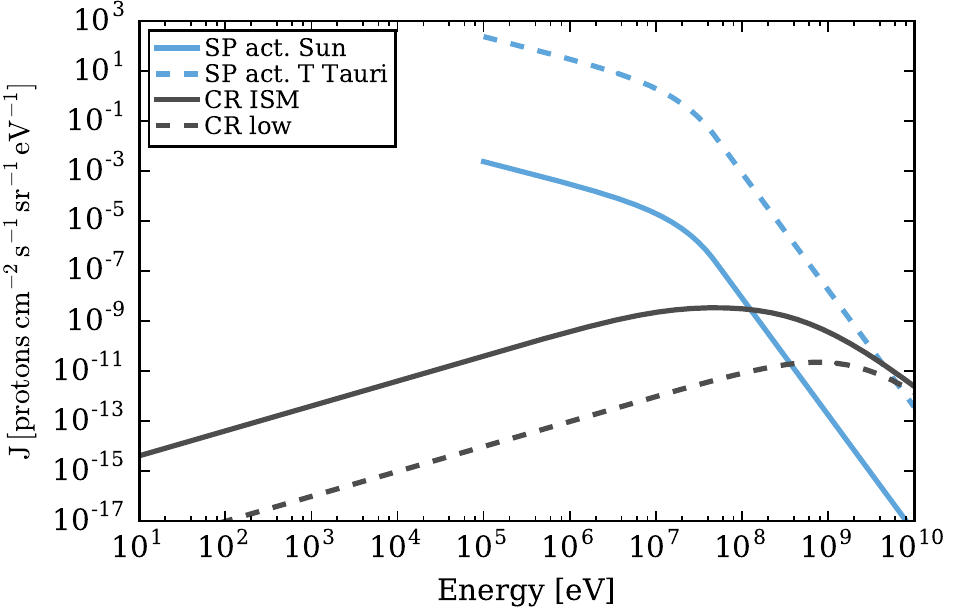}}
\caption{Stellar energetic proton (SP) and cosmic-ray (CR) input spectra. 
The blue solid and dashed lines show the active Sun and active T~Tauri SP spectrum, respectively. The black solid line shows the ``LIS  W98'' CR spectrum from \citet{Webber1998c} and the dashed black line the attenuated ``Solar Max'' CR spectrum from \citet{Cleeves2013}.}
\label{fig:spectra}
\end{figure}
\citet{Reedy2012d} reported proton fluxes of the contemporary Sun for 5 solar cycles. Typical values for the cycle averaged fluxes at a distance of $1\,\mathrm{au}$ are $f_\mathrm{p}(E_\mathrm{p}>10\,\mathrm{MeV})=59-213\,\mathrm{protons\,cm^{-2}\,s^{-1}}$. For the \emph{Event1} spectrum in Fig.~\ref{fig:spectrasun} we get $f_\mathrm{p}(E_\mathrm{p}>10\,\mathrm{MeV})=151\,\mathrm{protons\,cm^{-2}\,s^{-1}}$ at 1~au, very similar to the reported values of \citet{Reedy2012d}.   

Here we use only the \emph{Event1} spectrum and call this spectrum the ``active Sun'' spectrum. We simply scale the active Sun spectrum by a factor of $\approx10^5$, as proposed by \citet{Feigelson2002a}, to get a typical ``active T~Tauri'' spectrum. The resulting SP flux of $f_\mathrm{p}(E_\mathrm{p}>10\,\mathrm{MeV})=1.51\times10^7\,\mathrm{protons\,cm^{-2}\,s^{-1}}$ is consistent with the value of $\approx10^7\,\mathrm{protons\,cm^{-2}\,s^{-1}}$ derived by \citet{Feigelson2002a} for young solar analogs in the Orion Nebula cluster. The two SP spectra are shown in Fig.~\ref{fig:spectra}, where we also show two cases of Galactic cosmic-ray spectra for comparison (see~Sect.~\ref{sec:galacticCR}).

For comparison we also present models applying the same approach for the treatment of SPs as \citet{Turner2009bb}. For their model they assumed that SPs behave very similar to Galactic CRs (e.g. particle energies). The details of the Turner model are discussed in Appendix~\ref{sec:turnermodel}.

It is not clear if SPs actually reach the disk (see \citealt{Feigelson2002a} for a discussion). However, as we are interested in the possible impact of SPs on the chemistry, we simply assume that all SPs reach the disk. We further discuss this assumption in Sect.~\ref{sec:future}.

\subsection{Stellar particle transport and ionization rate}
\label{sec:sp}
Energetic particles hitting the disk interact with its gas and dust contents. 
Although the interaction with the solids is relevant for the production of SLR we are here only interested in the interaction with the gas. Dust only plays a minor role in the actual attenuation of particles as only $\approx1\%$ of the mass in protoplanetary disk is in solids (see also \citealt{Trappitsch2015}). Very similar to Galactic cosmic rays, SPs mainly ionize the gaseous medium (i.e. molecular hydrogen). Energetic particles interact multiple times and ionize many atoms/molecules on their way until they eventually have lost their energy completely. This complex process requires detailed particle transport models. 

To model the transport of energetic particles through the disk gas, we use the continuous slowing down approximation, which assumes that particles lose an infinitesimal fraction of their energy during propagation \citep{Takayanagi1973}. We use the results obtained by \citet{Padovani2009,Padovani2013c} who compute the propagation of CRs in a 1D slab, taking all the relevant energy loss processes into account. They give a useful fitting formula for the ionization rate as a function of the column density of molecular hydrogen. 

In order to apply the results of this 1D transport model, we assume that SPs travel along straight lines (e.g., no scattering due to their high energies) and that they originate from a point source (the star). We also neglect the effect of magnetic fields that could increase or decrease the ionization rate depending on their configuration (\citealt{Padovani2011c,Padovani2013}, see also Sect.~\ref{sec:mangeticfields}).

From the detailed 1D particle transport model we derive a simple fitting formulae for the two SP spectra considered here. The SP ionization rate $\zeta_\mathrm{SP}$ for molecular hydrogen as a function of the total hydrogen column density $N_\mathrm{<H>}=N_\mathrm{H}+2\times N_\mathrm{H_2}$ is given by
\begin{equation}
\zeta_\mathrm{SP}(N_\mathrm{<H>})=\left[\frac{1}{\zeta_\mathrm{L}\left(\frac{N_\mathrm{<H>}}{10^{20}\,\mathrm{cm^{-2}}}\right)^a} +\frac{1}{\zeta_\mathrm{H}\left(\frac{N_\mathrm{<H>}}{10^{20}\,\mathrm{cm^{-2}}}\right)^b}\right]^{-1}\;\;\mathrm{[s^{-1}]},
\label{eqn:strcfit}
\end{equation}
and for $N_{\mathrm{<H>}}> N_\mathrm{E}$ by
\begin{equation}
\zeta_\mathrm{SP,E}(N_\mathrm{<H>})=\zeta_\mathrm{SP}(N_\mathrm{<H>})\times\exp\left[-\left(\frac{N_\mathrm{<H>}}{N_\mathrm{E}}-1.0\right)\right]\;\;\mathrm{[s^{-1}]}. 
\label{eqn:strcfitH}
\end{equation}
The two power laws in Eq.~(\ref{eqn:strcfit}) are a consequence of the shape of the SP input spectra (see Fig.~\ref{fig:spectra}). The two parts of Eq.~(\ref{eqn:strcfit}) account for the ionization rate at low ($\zeta_\mathrm{L}$) and high $(\zeta_\mathrm{H}$) column densities. Equation~(\ref{eqn:strcfitH}) accounts for the exponential drop of the SP ionization rate starting at a certain column density given by $N_\mathrm{E}$ (i.e. similar to CRs). For the two SP spectra considered here $N_\mathrm{E}=2.5\times10^{25}\,\mathrm{cm^{-2}}$. The other fitting parameters $\zeta_\mathrm{L}$, $\zeta_\mathrm{H}$, $a$ and $b$ are given in Table~\ref{table:stcrfit}.

\begin{table}
\caption{Fitting parameters for the stellar particle ionization rate for the two different input spectra.}
\label{table:stcrfit}
\centering
\begin{tabular}{l c c c c}
\hline\hline
Name & $\zeta_\mathrm{L}$ & $a$ & $\zeta_\mathrm{H}$& $b$  \\
& $\mathrm{(s^{-1})}$ & & $\mathrm{(s^{-1})}$ & \\
\hline
SP active Sun   & $1.06(-12)$\tablefootmark{a} & $-0.61$ & $8.34(-7)$ & $-2.61$ \\
SP active T Tauri & $1.06(-7)$ & $-0.61$ & $8.34(-2)$ & $-2.61$ \\
\hline
\end{tabular}
\tablefoot{The values provided are for an unattenuated SP ionization rate at $1\,\mathrm{au}$ distance from the star (Equations~(\ref{eqn:strcfit}) and (\ref{eqn:strcfitH})).
\tablefoottext{a}{x(y) means $x\times10^y$}
}
\end{table}
Equations~(\ref{eqn:strcfit}) and (\ref{eqn:strcfitH}) provide the unattenuated SP ionization rate at 1~au distance from the star (i.e. for a SP flux at 1~au). To account for geometric dilution we scale $\zeta_\mathrm{SP}(N_\mathrm{<H>})$ by $1/r^2$ at every point in the disk ($r$ is the distance to the star in au). For the chemistry we simply add $\zeta_\mathrm{SP}$ to the ionization rate for Galactic cosmic rays $\zeta_\mathrm{CR}$ (see Sect.~\ref{sec:galacticCR}).
\subsection{Other ionization sources}
\label{sec:otherionsources}
To investigate the impact of SPs on the disk ionization structure also other ionization sources common to T~Tauri stars have to be considered. Besides SPs our model includes stellar UV and \mbox{X-ray} radiation, interstellar UV radiation and Galactic cosmic rays (CRs). However, most relevant for our study are the high energy ionization sources capable of ionizing molecular hydrogen: SPs, X-rays and CRs.
\subsubsection{X-rays}
\label{sec:xrays}
To model the stellar \mbox{X-ray} spectrum we use an approximation for an isothermal bremsstrahlung spectrum \citep{Glassgold1997a,Aresu2011}
\begin{equation}
F(E)\propto \frac{1}{E}\exp({-E/kT_\mathrm{X}}),
\end{equation}
where $E$ is the photon energy (here in the range of $0.1$ to $20\,\mathrm{keV}$), $k$ the Boltzmann constant and $T_\mathrm{X}$ is the plasma temperature. This spectrum is scaled to a given total \mbox{X-ray} luminosity $L_\mathrm{{X}}$ ($0.3\le E \le 10\,\mathrm{keV}$ e.g. \citealt{Gudel2010a}).

Due to the activity of the stars (e.g. flares) \mbox{X-ray} radiation of young stars is variable. We account for this in a simple way by including a spectrum with a \mbox{X-ray} luminosity and temperature representative for a typical T~Tauri star, and a spectrum which represents a flaring spectrum (more activity) with higher luminosity and a harder (hotter) radiation. 
However, we actually ignore the variability and assume time averaged \mbox{X-ray} fluxes. 
According to \citet{Ilgner2006f} this is a reasonable assumption as typically the recombination timescales in the disk are longer than the flaring period and large parts of the disk ($r\gtrsim 2\,\mathrm{au}$ in their model) respond to an enhanced average \mbox{X-ray} luminosity.  

\begin{figure}
\resizebox{\hsize}{!}{\includegraphics{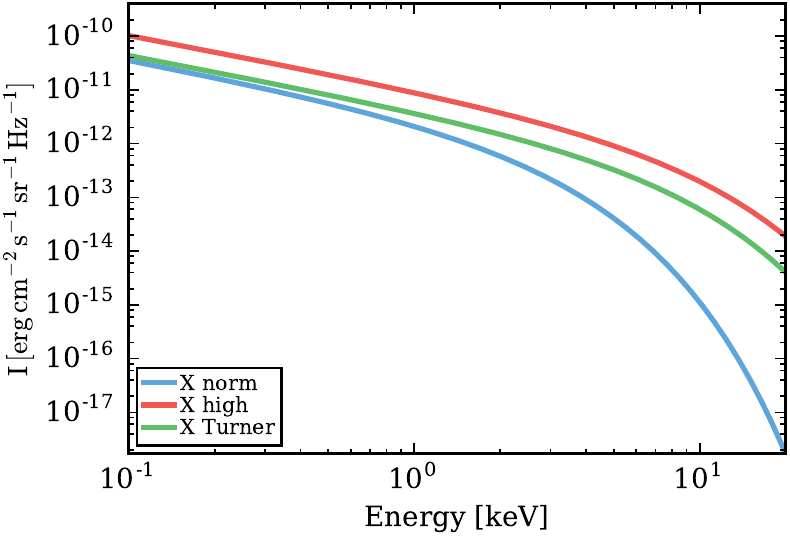}}
\caption{Stellar X-ray input spectra (see Table~\ref{table:xparam}).}
\label{fig:xray_spec}
\end{figure}

Additionally we also include the stellar \mbox{X-ray} properties used in \citet{Turner2009bb} for the Turner models (Appendix~\ref{sec:turnermodel}). The parameters for the various \mbox{X-ray} input spectra are given in Table~\ref{table:xparam} and the spectra are shown in Fig.~\ref{fig:xray_spec}.

To derive the \mbox{X-ray} ionization rate $\zeta_\mathrm{X}$ we use \mbox{X-ray} radiative transfer including scattering and a detailed treatment of \mbox{X-ray} chemistry \citep{Aresu2011,Meijerink2012a}. For more details on the new \mbox{X-ray} radiative transfer module in P{\tiny RO}D{\tiny I}M{\tiny O} see Appendix~\ref{sec:app_xrt}.
\begin{table}
\caption{Parameters for the stellar X-ray input spectra (see Fig.~\ref{fig:xray_spec}).}
\label{table:xparam}
\centering
\begin{tabular}{l c c c }
\hline\hline
Name & Spectrum & $L_\mathrm{X}$ & $T_\mathrm{X}$ \\
& & $\mathrm{(erg\,s^{-1})}$ &  (K) \\
\hline
X norm & typical T~Tauri & 1(30) & 2(7) \\
X high & flared T~Tauri  & 5(30) & 7(7)\\
X Turner & \citet{Turner2009bb} & 2(30) & 5.8(7) \\
\hline
\end{tabular}
\end{table}
\subsubsection{Galactic cosmic rays}
\label{sec:galacticCR}
Protoplanetary disks are exposed to Galactic cosmic rays (CR). Differently to SPs and X-rays, CRs are not of stellar origin and hit the disk isotropically. \citet{Cleeves2013} proposed that for T~Tauri disks the actual CR ionization rate might be much lower compared to the interstellar medium (ISM) due to modulation of the impinging CRs by the heliosphere (``\mbox{T-Tauriosphere}''). 

From modeling molecular ion observations of the \object{TW Hya} disk, \citet{Cleeves2015a} derived an upper limit for the total H$_2$ ionization rate of $\mathrm{\zeta\lesssim10^{-19}\,s^{-1}}$. This upper limit applies for all ionization sources including SLRs \citep[e.g.][]{Umebayashi2009}. SLR ionization is a potentially important ionization source in the midplane of disks. However, similar to \citet{Cleeves2015a} we model the low ionization rate scenario by reducing the CR ionization rate and do not explicitly treat SLR ionization in the models presented here.  

We consider two different CR input spectra, the canonical local ISM CR spectrum \citep{Webber1998c} and a modulated CR spectrum which accounts for the exclusion of CRs by the ''T-Tauriosphere'' \citep{Cleeves2013,Cleeves2015a}. For simplicity we call these two spectra ``ISM CR'' and ``low CR'', respectively. To calculate the CR ionization rate $\zeta_\mathrm{CR}$ in the disk we apply the fitting formulae provided by \citet{Padovani2013c} and \citet{Cleeves2013}: 
\begin{equation}
\label{eqn:crfit}
\zeta_\mathrm{CR}(N_\mathrm{<H>})=\frac{\zeta_\mathrm{l}\,\zeta_\mathrm{h}}{\zeta_\mathrm{h}[N_\mathrm{<H>}/10^{20}\,\mathrm{cm^{-2}}]^a+\zeta_\mathrm{l}[\exp(\Sigma/\Sigma_0)-1]}.
\end{equation}
For simplicity we assume that CRs enter the disk perpendicular to the disk surface. Therefore we use the disk vertical hydrogen column density $N_\mathrm{<H>,ver}$ and surface density $\Sigma_\mathrm{ver}$ for Eq.~(\ref{eqn:crfit}) to calculate $\zeta_\mathrm{CR}$ at every point in the disk. The fitting parameters $\zeta_\mathrm{l}$, $\zeta_\mathrm{h}$, $\Sigma_0$ and $a$ for the two CR input spectra are given in Table~\ref{table:crparam} (see \citealt{Padovani2009,Padovani2013c} for details). The typical resulting H$_2$ ionization rates in the disk are $\mathrm{\zeta_{CR}\approx2\times10^{-17}\,s^{-1}}$ for the ISM CR spectrum and $\mathrm{\zeta_{CR}\approx2\times10^{-19}\,s^{-1}}$ for the low CR spectrum (see Sect.~\ref{sec:resionrate} and \ref{sec:diskionrates}).
\begin{table}
\caption{Fitting parameters for the cosmic-ray ionization rate. 
}
\label{table:crparam}
\centering
\begin{tabular}{l c c c c}
\hline\hline
Name & $\zeta_\mathrm{l}$ & $\zeta_\mathrm{h}$& $\mathrm{\Sigma_{0}}$ & $a$  \\
& $\mathrm{(s^{-1})}$ & $\mathrm{(s^{-1})}$ & $\mathrm{(g\,cm^{-2})}$ & \\  
\hline
ISM CR \tablefootmark{a} & $2(-17)$ & $2.6(-18)$ & $244$ & $0.021$ \\
low CR \tablefootmark{b}& $2(-19)$ & $8.0(-19)$ & $230$ & $-0.01$ \\
\hline
\end{tabular}
\tablefoot{
\tablefoottext{a}{ISM W98 spectrum \citep{Padovani2009,Padovani2013c}}\\
\tablefoottext{b}{modulated ``Solar Max'' spectrum \citep{Cleeves2013}}
}
\end{table}
\begin{figure*}
\centering
\resizebox{\hsize}{!}{\includegraphics{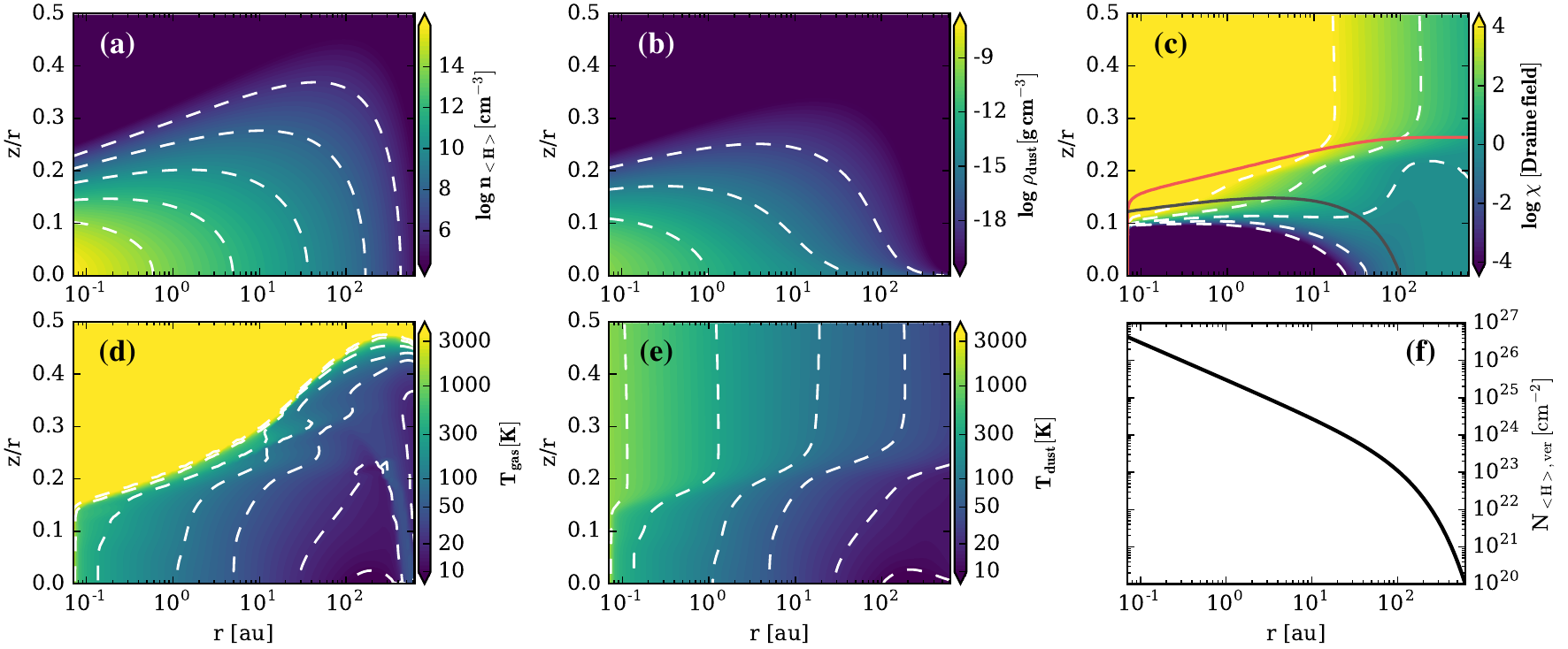}}
\caption{Two dimensional structure of the reference disk model CI\_XN. The height of the disk $z$ is scaled by the radius ($z/r$). From top left to the bottom right: \emph{(a)} gas number density $n_\mathrm{<H>}$, \emph{(b)} dust density $\rho_\mathrm{dust}$ (note the dust settling), \emph{(c)} FUV radiation field $\chi$ in units of the ISM Draine field, \emph{(d)} gas temperature $T_\mathrm{gas}$, \emph{(e)} dust temperature $T_\mathrm{dust}$ and \emph{(f)} the vertical hydrogen column density $N_\mathrm{<H>,ver}$ versus radius. The white dashed contour lines in each contour plot correspond to the levels shown in the respective colorbar. The black (red) solid contour in panel \emph{(c)} indicate a vertical (radial) visual extinction equal to unity.}
\label{fig:diskstruc}
\end{figure*}
\subsection{Disk model}
\label{sec:diskmodel}
To model the disk we use the radiation thermo-chemical disk code P{\tiny RO}D{\tiny I}M{\tiny O} \citep{Woitke2009a,Kamp2010,Thi2011,Woitke2016}. P{\tiny RO}D{\tiny I}M{\tiny O} solves the wavelength dependent continuum radiative transfer which provides the disk dust temperature and the local radiation field. The gas temperature (heating/cooling balance) is determined consistently with the chemical abundances. The chemical network includes 235 different species and 3143 chemical reactions (see Appendix~\ref{sec:chemistry} for more details).

We use a disk model representing the main properties of a disk around a typical T~Tauri star. The stellar properties and the disk structure of this model are identical to the so called ``reference model'' presented in \citet{Woitke2016}. Here we only provide a brief overview of the disk model and refer the reader to \citet{Woitke2016} for details. 

In Fig.~\ref{fig:diskstruc} we show the gas number density, dust density, the local far-UV (FUV) radiation field, gas temperature, dust temperature and the vertical hydrogen column density for the reference model (model CI\_XN, see Sect.~\ref{sec:modelgrid} and Table~\ref{table:models}). All relevant parameters of the disk model are given in Table~\ref{table:discmodel}.

\begin{table}
\caption{Main fixed parameters of the disk model.}
\label{table:discmodel}
\centering
\begin{tabular}{l|c|c}
\hline\hline
Quantity & Symbol & Value  \\
\hline
stellar mass                          & $M_\mathrm{*}$                    & $0.7~\mathrm{M_{\sun}}$\\
stellar effective temp.               & $T_{\mathrm{*}}$                  & 4000~K\\
stellar luminosity                    & $L_{\mathrm{*}}$                  & $1.0~\mathrm{L_{\sun}}$\\
FUV excess                            & $L_{\mathrm{FUV}}/L_{\mathrm{*}}$ & 0.01\\
FUV power law index                   & $p_{\mathrm{UV}}$                 & 1.3\\
\hline
strength of interst. FUV\tablefootmark{a}              & $\chi^\mathrm{ISM}$               & 1\\
\hline
disk gas mass                         & $M_{\mathrm{disk}}$               & $0.01~\mathrm{M_{\sun}}$\\
dust/gas mass ratio                   & $d/g$                             & 0.01\\
inner disk radius                     & $R_{\mathrm{in}}$                 & 0.07~au\\
tapering-off radius                   & $R_{\mathrm{tap}}$                & 100~au\\
column density power ind.             & $\gamma$                        & 1.0\\

reference scale height                & $H(100\;\mathrm{au})$             & 10 au\\
flaring power index                   & $\beta$                           & 1.15\\
\hline
min. dust particle radius             & $a_\mathrm{min}$                  & $\mathrm{0.05~\mu m}$\\
max. dust particle radius             & $a_\mathrm{max}$                  & 3 mm\\
dust size dist. power index           & $a_\mathrm{pow}$                  & 3.5\\
turbulent mixing param.               & $\alpha_{\mathrm{settle}}$        & $10^{-3}$\\
max. hollow volume ratio\tablefootmark{b}              & $V_{\mathrm{hollow,max}}$         & 0.8\\
dust composition                      & {\small Mg$_{0.7}$Fe$_{0.3}$SiO$_3$}  & 60\%\\
(volume fractions)                    & {\small amorph. carbon}                    & 15\%\\
& {\small porosity}                          & 25\%\\
\hline
\end{tabular}
\tablefoot{
If not noted otherwise, these parameters are kept fixed for all our models presented in this work. For more details on the parameter definitions see \citet{Woitke2009a,Woitke2011,Woitke2016}.\\
\tablefoottext{a}{$\chi^\mathrm{ISM}$ is given in units of the Draine field \citep{Draine1996b,Woitke2009a}.}\\
\tablefoottext{b}{We use distributed hollow spheres for the dust opacity calculations \citep{Min2005,Min2016ax}.}
}
\end{table}
Based on the similarity solution for viscous accretion disks, we use an axisymmetric flared gas density structure with a Gaussian vertical profile and a powerlaw with a tapered outer edge for the radial column density profile \citep[e.g.][]{Lynden-Bell1974b,Andrews2009a}. The vertical scale height as a function of radius is expressed by a simple powerlaw. The disk has a total mass of $0.01\,\mathrm{M_\sun}$ and extends from $0.07\,\mathrm{au}$ (the dust sublimation radius) to $620\,\mathrm{au}$ where the total vertical hydrogen column density reaches $N\mathrm{_{<H>,ver}\approx10^{20}\,cm^{-2}}$ (panel (f) in Fig.~\ref{fig:diskstruc}).

For the dust density distribution we assume a dust to gas mass ratio of $d/g=0.01$. There is observational evidence for dust growth and settling in protoplanetary disks \citep[e.g.][]{Williams2011,Dullemond2004b}. To account for dust growth we assume a power law dust size distribution $f(a)=a^{-3.5}$ with a minimum and maximum grain radius of $a_\mathrm{min}=0.05\,\mathrm{\mu m}$ and $a_\mathrm{max}=3000\,\mathrm{\mu m}$. For dust settling we apply the method of \citet{Dubrulle1995} with a turbulent mixing parameter of $10^{-3}$.

The irradiation of the disk by the star is important for the temperature and the chemical composition of the disk. For the photospheric emission of the star we use PHOENIX stellar atmosphere models \citep{Brott2005a}. We consider a $\rm{0.7\,M_{\sun}}$ star with an effective temperature of 4000$\,$K and a luminosity of 1$\,\rm{L_\sun}$. In addition to the photospheric emission, T~Tauri stars commonly show far ultra-violet (FUV) excess \citep[e.g.][]{France2014b} due to accretion shocks and strong \mbox{X-ray} emission \citep[e.g.][]{Guedel2009g}. For the excess FUV emission, we use a simple power law spectrum with a total integrated FUV luminosity of $L_{\rm{FUV}}=0.01\,\mathrm{L_*}$, in the \mbox{wavelength interval [91.2$\,$nm, 250$\,$nm]}. The details for the stellar \mbox{X-ray} properties were already discussed in Sect.~\ref{sec:xrays}.

We mainly use the the molecules HCO$^+$ and N$_2$H$^+$ to study the impact of SP ionization. To verify if our model gives reasonable results concerning HCO$^+$ and N$_2$H$^+$ abundances we compare the modelled fluxes for the $\mathrm{J=3-2}$ transition of $\mathrm{HCO^+}$ and $\mathrm{N_2H^+}$ with the observational sample of \citet{Oberg2010c,Oberg2011a} finding a good agreement (for details see Appendix~\ref{sec:cobs}).
\subsection{Model series}
\label{sec:modelgrid}
It is likely that the different ionization sources are correlated. As already discussed the SP flux of young stars is actually derived from their stellar \mbox{X-ray} properties \mbox{\citep{Lee1998mq,Feigelson2002a}}. Also the CR ionization rate might be anti-correlated with the activity of the star \citep{Cleeves2013}. However, these possible correlations are not well understood. We therefore run a series of full disk models using the already described \mbox{X-ray}, SP and CR spectra as inputs and also include the Turner SP model (Appendix~\ref{sec:turnermodel}). We do not discuss any model with the active Sun SP spectrum as in this case SPs do not have a significant impact on the disk chemical structure (see Sec.~\ref{sec:resionrate}). An overview of all presented models is given in Table~\ref{table:models}.
\begin{table}
\caption{Model series.}
\label{table:models}
\centering
\begin{tabular}{l l l l }
\hline\hline
Name & X-rays &  Stellar particles & Cosmic rays \\
\hline
CI\_XN      & normal\tablefootmark{a} & $-$ & ISM \tablefootmark{d} \\
CI\_XH      & high\tablefootmark{b}  & $-$ & ISM \\
CI\_XN\_SP  & normal & active T~Tauri & ISM \\
CI\_XH\_SP  & high   & active T~Tauri & ISM \\
CI\_T       & Turner\tablefootmark{c} & Turner & ISM \\
CL\_XN      & normal & $-$ & low \tablefootmark{e} \\
CL\_XH      & high   & $-$ & low \\
CL\_XN\_SP  & normal & active T~Tauri & low \\
CL\_XH\_SP  & high   & active T~Tauri & low \\
CL\_T       & Turner & Turner & low \\
\hline
\end{tabular}
\tablefoot{
In the model names CI (CL) stands for ISM (low) CR ionization rates, XN (XH) for normal (high) X-ray luminosities, SP for stellar particles and T for Turner model. \\
\tablefoottext{a}{X-ray luminosity $L_\mathrm{X}=10^{30}\,\mathrm{erg\,s^{-1}}$.}\\
\tablefoottext{b}{$L_\mathrm{X}=5\times10^{30}\,\mathrm{erg\,s^{-1}}$.}
\tablefoottext{c}{$L_\mathrm{X}=2\times10^{30}\,\mathrm{erg\,s^{-1}}$.}\\
\tablefoottext{d}{CR ionization rate $\zeta_\mathrm{CR}\approx2\times10^{-17}\,\mathrm{s^{-1}}$.}
\tablefoottext{e}{$\zeta_\mathrm{CR}\approx2\times10^{-19}\,\mathrm{s^{-1}}$.}
}
\end{table}
\section{Results}
\label{sec:results}
\subsection{Ionization rates as a function of column density}
\label{sec:resionrate}
\begin{figure}
\resizebox{\hsize}{!}{\includegraphics{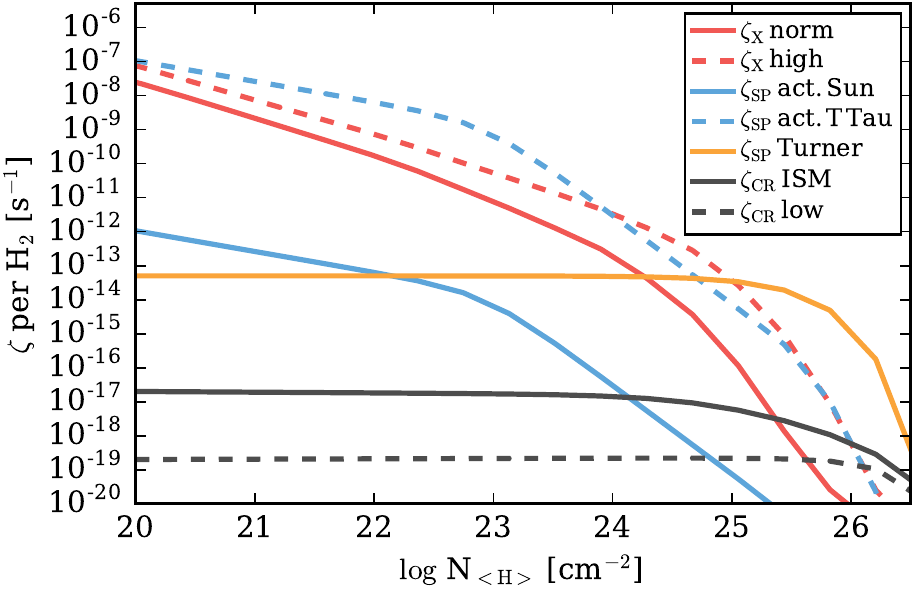}}
\caption{SP, CR and X-ray ionization rates $\zeta$ as a function of hydrogen column density $N\mathrm{_{<H>}}$.}
\label{fig:ionR}
\end{figure}

Before we discuss our results for the full disk model, we compare the SP, \mbox{X-ray} and CR ionization rates as a function of the total hydrogen column density $N\mathrm{_{<H>}}$ ($N_\mathrm{<H>}=N_\mathrm{H}+2N_\mathrm{H_2}$). Fig.~\ref{fig:ionR} shows such a comparison for our different input spectra discussed in Sections \ref{sec:sp} and \ref{sec:otherionsources}.

From Fig.~\ref{fig:ionR} it becomes clear that for a SP flux comparable to our Sun (active Sun spectrum) SP ionization cannot compete with \mbox{X-ray}  ionization assuming typical T~Tauri \mbox{X-ray} luminosities. However, for the active T~Tauri SP spectrum SP ionization becomes comparable to \mbox{X-ray} ionization or even dominates for $N\mathrm{_{<H>}}\lesssim10^{24}-10^{25}\,\mathrm{cm^{-2}}$. For $N_\mathrm{H}\lesssim 10^{23}\,\mathrm{cm^{-2}}$ $\zeta_\mathrm{SP}$ is determined by the particles with $E_\mathrm{p}\lesssim 5\times10^7\,\mathrm{eV}$ whereas higher energy particles dominate for $N\mathrm{_{<H>}}\gtrsim 10^{23}\,\mathrm{cm^{-2}}$. The kink at $N\mathrm{_{<H>}}\approx 2\times10^{25}\,\mathrm{cm^{-2}}$ is caused by the rapid attenuation of the SPs at high column densities. At this high column densities even the most energetic particles have lost most of their energy and the ionization rate drops exponentially.

For X-rays, Fig.~\ref{fig:ionR} shows the differences between the normal and high/harder \mbox{X-ray} spectrum. The \mbox{X-ray} ionization rates are higher for the high \mbox{X-ray} spectrum due to the higher \mbox{X-ray} luminosity. Additionally, the harder \mbox{X-ray} photons can penetrate to deeper layers but are also more efficiently scattered than lower energy \mbox{X-ray} photons. Compared to the normal \mbox{X-ray} case, the \mbox{X-ray} ionization rate increases by several orders of magnitude for $N\mathrm{_{<H>}}\gtrsim10^{24}\,\mathrm{cm^{-2}}$ for the high \mbox{X-ray} case.

Galactic cosmic rays are the most energetic ionization source. The peak in the particle energy distribution is around $10^8 - 10^9\,\mathrm{eV}$ (see~Fig.~\ref{fig:spectra}). As a consequence the CR ionization rate $\zeta_\mathrm{CR}$ stays mostly constant and only decrease for \mbox{$N\mathrm{_{<H>}}\gtrsim 10^{25}\,\mathrm{cm^{-2}}$}. Only for such high column densities CR particle absorption becomes efficient.

In the Turner model it is implicitly assumed that SPs have the same energy distribution than Galactic CRs (see Appendix~\ref{sec:turnermodel}). As a consequence the SP ionization rate in the Turner model is simply a scaled up version of the CR ionization rate. The slight differences to our model in CR attenuation is caused by the different methods used to calculate the SP/CR ionization rates; \citet{Turner2009bb} use the fitting formulae of \citet{Umebayashi2009}. Compared to our active T~Tauri SP spectrum $\zeta_\mathrm{SP}$ in the Turner model is larger for $N_\mathrm{<H>}>10^{25}\,\mathrm{cm^{-2}}$ but significantly lower at low column densities.
\subsection{Disk ionization rates}
\label{sec:diskionrates}
\begin{figure*}
\resizebox{\hsize}{!}{\includegraphics{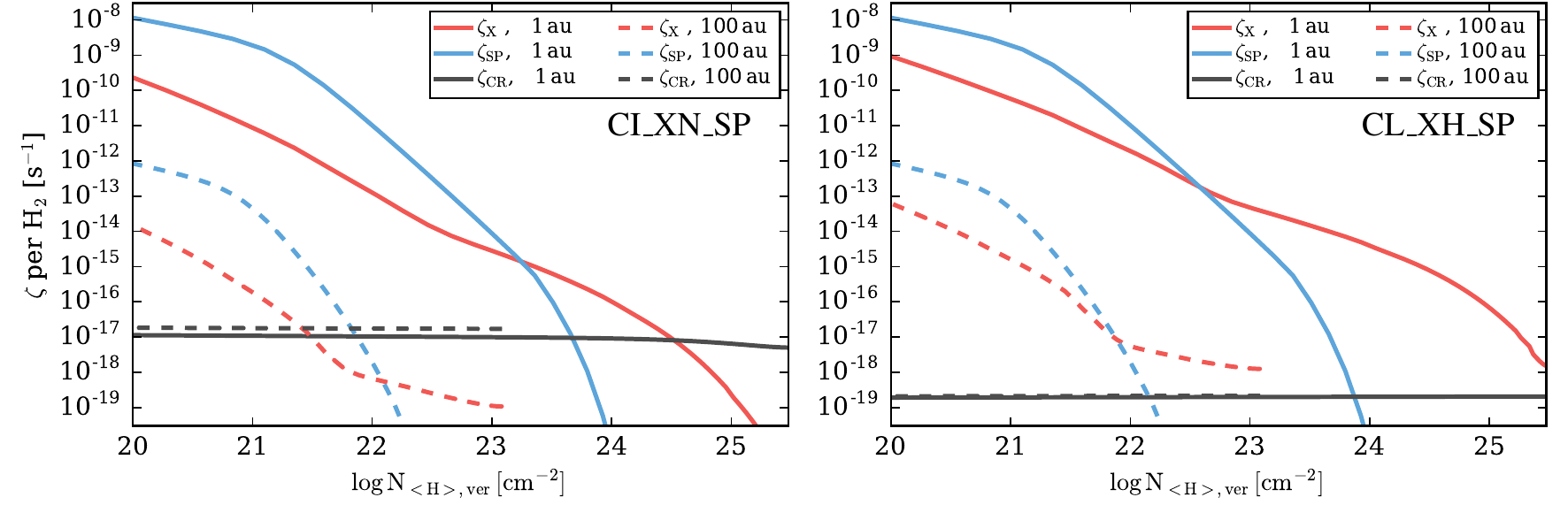}}
\caption{Ionization rates $\zeta$ as a function of vertical column density $N\mathrm{_{<H>,ver}}$ at radii of 1 and 100 au (solid and dashed lines respectively). 
The maximum values for $N_\mathrm{<H>,ver}$ at the midplane of the disk, are $N_\mathrm{<H>,ver}\approx4\times10^{25}\,\mathrm{cm^{-2}}$ and $N\mathrm{_{<H>,ver}}\approx2\times10^{23}\,\mathrm{cm^{-2}}$ at $1\,\mathrm{au}$ and $100\,\mathrm{au}$, respectively. Red lines are for X-rays, blue lines are for SPs and the black lines are for CRs. \emph{Left panel:} model CI\_XN\_SP with ISM CRs and normal X-rays; \emph{right panel:} model CL\_XH\_SP with low CRs and high X-rays.} 
\label{fig:ionR_disk2}
\end{figure*}
In Fig.~\ref{fig:ionR_disk2} we show the ionization rates as a function of the vertical hydrogen column density $N\mathrm{_{<H>,ver}}$ at two different radii of the disk. Shown are the models CI\_XN\_SP (ISM CR, normal X-rays; left panel) and CL\_XH\_SP (low CR, high X-rays; right panel). 

CRs are only significantly attenuated for \mbox{$N_\mathrm{<H>,ver}>10^{25}\,\mathrm{cm^{-2}}$} and $r\lesssim1\,\mathrm{au}$. For most of the disk, CRs provide a nearly constant ionization rate of $\zeta_\mathrm{CR}\approx2\times10^{-17}\,\mathrm{s^{-1}}$ for the ISM like and $\zeta_\mathrm{CR}\approx2\times10^{-19}\,\mathrm{s^{-1}}$ for the low CR spectrum.

X-rays are strongly attenuated as a function of height and radius (i.e. geometric dilution). However, due to scattering X-rays can become the dominant midplane ionization source for large regions of the disk. For ISM like CRs, CR ionization is the dominant midplane ionization source even in the high \mbox{X-ray} models. In the low CRs models X-rays are the dominant midplane ionization source for $r\lesssim100\,\mathrm{au}$ in the normal \mbox{X-ray} model and for all radii in the high \mbox{X-ray} model.

\begin{figure*}
\resizebox{\hsize}{!}{\includegraphics{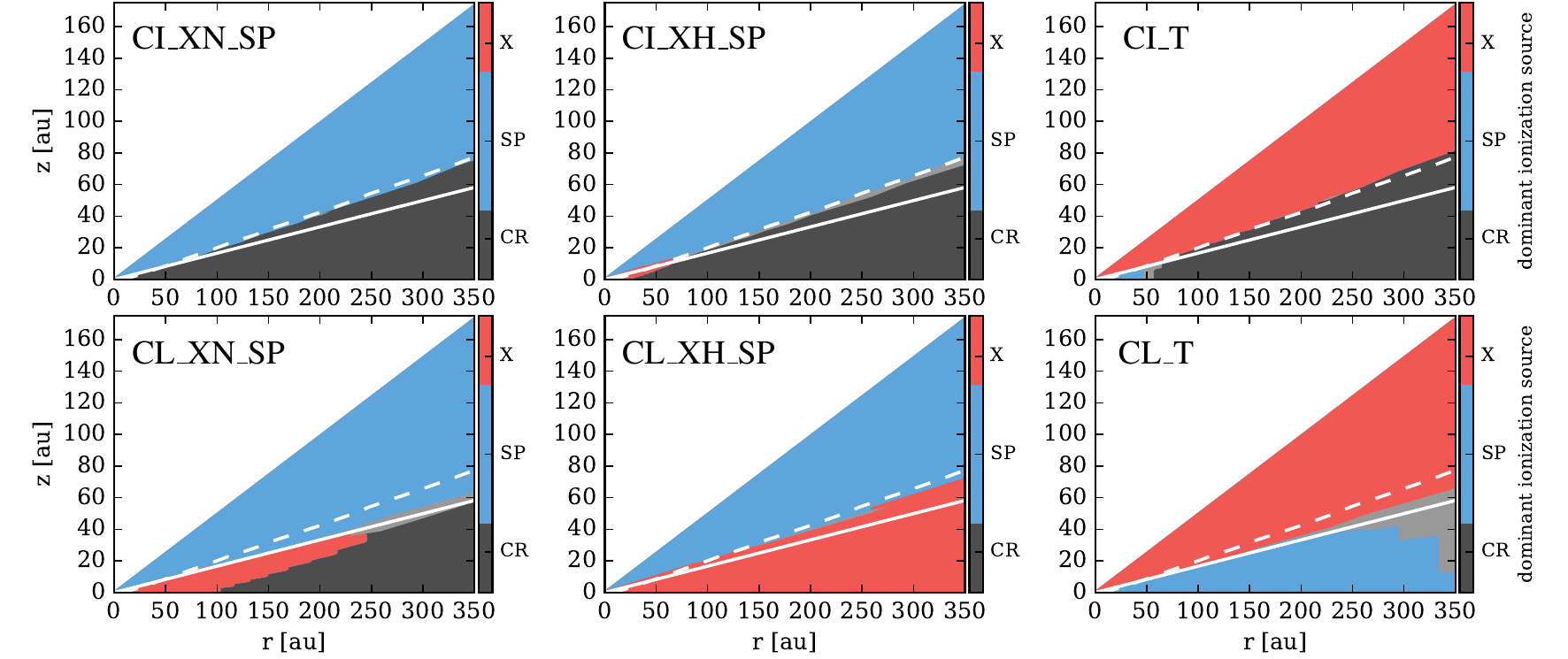}}
\caption{Dominant disk ionization source throughout the disk. An ionization source is dominant at a certain point in the disk if its value is higher than the sum of the two other ionization sources. A light gray area indicates a region without a dominant ionization source. The different possible ionization sources, X-rays, SPs and CRs are identified by the different colors (color bar). The white solid contour line shows $N\mathrm{_{<H>,rad}}=10^{25}\,\mathrm{cm^{-2}}$ the white dashed line shows the CO ice line. The model names are given in the top left of each panel. \emph{Top row:} models with ISM CR ionization rate (CI); \emph{bottom row:} models with low CR ionization rate (CL). \emph{First column:} normal \mbox{X-ray} models (XN); \emph{second column:} high \mbox{X-ray} models (XH); \emph{third column:} Turner models (T).}
\label{fig:ionR_dom}
\end{figure*}
Differently to X-rays, SPs are not scattered towards the midplane. Due to their high energies they propagate along straight lines (provided that the SPs are not shielded by magnetic fields, see Sect.~\ref{sec:mangeticfields}). As SPs are of stellar origin they penetrate the disk only along radial rays. The radial column densities close to the midplane of the disk are $N\mathrm{_{<H>,rad}}\gg10^{25}\,\mathrm{cm^{-2}}$ and therefore SPs are already strongly attenuated at the inner rim of the disk. From Fig.~\ref{fig:ionR_disk2} we see that the SP ionization rate $\zeta_\mathrm{SP}$ drops below $10^{-19}\,\mathrm{s^{-1}}$ for $N\mathrm{_{<H>,ver}}\gtrsim 10^{24}\,\mathrm{cm^{-2}}$ at $r=1\,\mathrm{au}$ and for $N\mathrm{_{<H>,ver}}\gtrsim 10^{22}\,\mathrm{cm^{-2}}$ at $r=100\,\mathrm{au}$. However, at higher layers ($N_\mathrm{<H>,ver}\lesssim 10^{22}-10^{23}\,\mathrm{cm^{-2}}$) SPs are the dominant ionization source even in the high \mbox{X-ray} models. Expressed in radial column densities: SPs are the dominant ionization source in disk regions with $N_\mathrm{<H>,rad}\lesssim10^{24}-10^{25}\,\mathrm{cm^{-2}}$.

In Fig.~\ref{fig:ionR_dom} we show the dominant ionization source at every point in the disk for all SP models. The first two columns show ``our'' models, the last column the Turner models. In our models, SPs are the dominant ionization source in the upper layers of the disk (above the white solid contour line for $N\mathrm{_{<H>,rad}}=10^{25}\,\mathrm{cm^{-2}}$), whereas in the midplane always CRs or X-rays dominate.

For the Turner model the picture is quite different. In their model SPs can penetrate the disk also vertically (Appendix~\ref{sec:turnermodel}) and reach higher vertical column densities before they are completely attenuated (Fig.~\ref{fig:ionR}). As a consequence SPs can become the dominant ionization source in the midplane of the disk (e.g. for the low CR case). In the upper layers always X-rays dominate as $\zeta_\mathrm{SP}<\zeta_\mathrm{X}$ for low column densities. In the Turner model $\zeta_\mathrm{SP}\lesssim10^{-13}\,\mathrm{s^{-1}}$ for $N\mathrm{_{<H>,rad}}<10^{25}\,\mathrm{cm^{-2}}$ which is orders of magnitudes lower than in our models. The reason for this is that in the Turner model SPs are simply a scaled version of ISM like CRs. The high $\zeta_\mathrm{SP}$ values in our model in the upper layers of the disk are caused by the high number of particles with energies $E_\mathrm{p}\lesssim 10^8\,\mathrm{eV}$, which are missing in the Turner model. 
\subsection{Impact on HCO$^+$ and N$_2$H$^+$}
\label{sec:HcopN2Hp}
The molecules HCO$^+$ and N$_2$H$^+$ are the two most observed molecular ions in disks \citep[e.g.][]{Dutrey2014a} and are commonly used to trace the ionization structure of disks \citep[e.g.][]{Dutrey2007a,Oberg2011f,Cleeves2015a}. Also \citet{Ceccarelli2014d} used these two molecules to trace SPs in a protostellar envelope.

\begin{figure*}
\resizebox{\hsize}{!}{\includegraphics{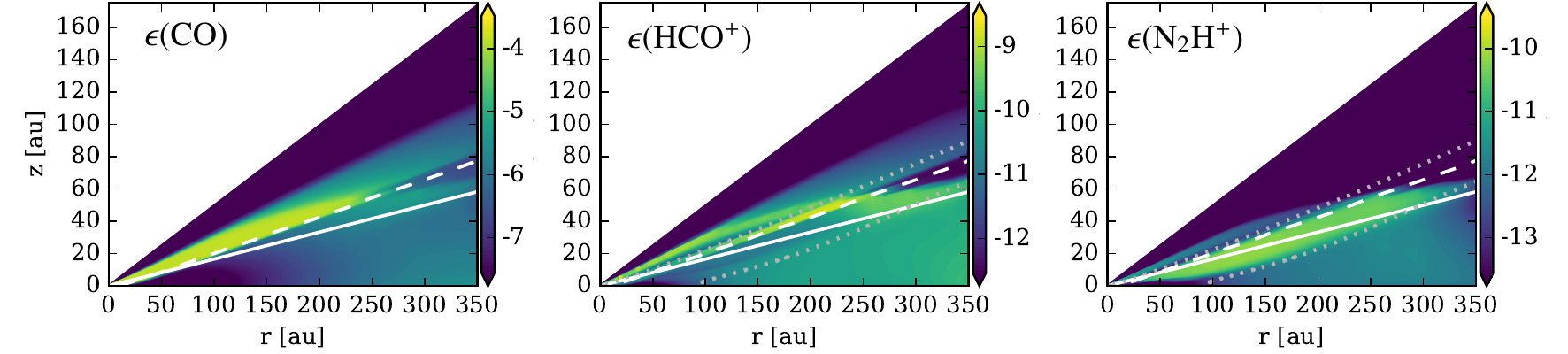}}
\caption{Abundances $\epsilon$(X) relative to hydrogen for CO, HCO$^+$ and N$_2$H$^+$ for the reference model CI\_XN. The white solid contour line shows $N_\mathrm{<H>,rad}=10^{25}\,\mathrm{cm^{-2}}$, the white dashed line shows the CO ice line. We call the regions above and below the CO ice line the warm and cold molecular layer, respectively.  
The dotted iso-contours show where the \mbox{X-ray} ionization rate is equal to the ISM CR ($\mathrm{\zeta_{CR}=2\times10^{-17}\,s^{-1}}$) and equal to the low CR ($\mathrm{\zeta_{CR}=2\times10^{-19}\,s^{-1}}$) ionization rate, respectively.
}
\label{fig:abun}
\end{figure*}
The main formation path of HCO$^+$ and N$_2$H$^+$ is the ion-neutral reaction of H$_3^+$ with their parent molecules CO and N$_2$, respectively. H$_3^+$ is created by ionization of H$_2$ by CRs, X-rays and in our model additionally by SPs. The main destruction pathway for HCO$^+$ and N$_2$H$^+$ is via dissociative recombination with free electrons. 

The chemistry of HCO$^+$ and N$_2$H$^+$ is linked to the freeze-out of CO. To form HCO$^+$, gas phase CO is required, whereas N$_2$H$^+$ is efficiently destroyed by CO \citep[e.g.][]{Aikawa2015}. Consequently the N$_2$H$^+$ abundance peaks in regions where CO is depleted and N$_2$, the precursor of N$_2$H$^+$, is still in the gas phase. The result of this chemical interaction is a vertically layered chemical structure for HCO$^+$ and N$_2$H$^+$ (see Fig.~\ref{fig:abun}). For further details on the HCO$^+$ and N$_2$H$^+$ chemistry see Appendix~\ref{sec:chemistry}, where we also list the main formation/destruction pathways for HCO$^+$ and N$_2$H$^+$ (Table~\ref{table:ratecoeff}).
\subsubsection{Abundance structure}
\label{sec:abunstruc}
In the following we describe details of the molecular abundance structure that are relevant for the presentation of our results for our reference model CI\_XN. The abundance $\epsilon$ of a molecule X is given by $\epsilon(\mathrm{X})=n_\mathrm{X}/n_\mathrm{<H>}$, where $n_\mathrm{X}$ is the number density of the respective molecule and $n_\mathrm{<H>}=n_\mathrm{H}+2\,n_\mathrm{H_2}$ is the total hydrogen number density. Fig.~\ref{fig:abun} shows the resulting abundance structure for CO, HCO$^+$ and N$_2$H$^+$ for the CI\_XN model.  

We define the location of the CO ice line where the CO gas phase abundance is equal to the CO ice-phase abundance (white dashed line in Fig.~\ref{fig:abun}). The CO ice line is located at dust temperatures in the range $T_\mathrm{d}\approx23-32\,\mathrm{K}$ (density dependence of the adsorption/desorption equilibrium; e.g.~\citealt{Furuya2014g}). The radial CO ice line in the midplane ($z=0\,\mathrm{au}$) is at $r\approx12\,\mathrm{au}$ and $T_\mathrm{d}\approx32\,\mathrm{K}$. At $r\approx50\,\mathrm{au}$ the vertical CO ice line is at $z\approx8.5\,\mathrm{au}$ ($z/r\approx0.17$) and $T_\mathrm{d}\approx26\,\mathrm{K}$. Inside/above the CO ice line $\mathrm{\epsilon(CO)}\approx 10^{-4}$. Outside/below the CO ice line $\mathrm{\epsilon(CO)}$ rapidly drops to values $\lesssim10^{-6}$. In regions where non-thermal desorption processes are efficient ($r\gtrsim 150\,\mathrm{au}$) $\mathrm{\epsilon(CO)}\approx10^{-6}$ down to the midplane. For the regions inside/above and outside/below the CO ice line we use the terms warm and cold molecular layer, respectively.

There are two main reservoirs for HCO$^+$, one in the warm molecular layer above the CO ice line and one in the outer disk ($r\gtrsim150\,\mathrm{au}$) below the CO ice line where non-thermal desorption becomes efficient. In the warm molecular layer, the ionization fraction $\mathrm{\epsilon(e^-)}\approx10^{-7}$ is dominated by sulphur as it is ionized by UV radiation (e.g. \citealt{Teague2015b}, see also Sect.~\ref{sec:metalabun}). Those free electrons efficiently destroy molecular ions via dissociative recombination. This causes a dip in the vertical HCO$^+$ abundance structure within the warm molecular layer with $\mathrm{\epsilon(HCO^+)\approx10^{-12}-10^{-11}}$, whereas at the top and the bottom of the warm molecular layer $\mathrm{\epsilon(HCO^+)}$ reaches values of $\approx10^{-10}-10^{-9}$. The peak in the top layer is mainly caused by the high \mbox{X-ray} ionization rate for H$_2$ ($\mathrm{\zeta_{X}\gtrsim10^{-12}\,s^{-1}}$). At the bottom of the warm molecular layer more HCO$^+$ survives. This region is already sufficiently shielded from UV radiation and the free electron abundance drops rapidly. In the second reservoir, below the CO ice line where non-thermal desorption is efficient $\mathrm{\epsilon(HCO^+)}\approx10^{-11}-10^{-10}$.   

The main N$_2$H$^+$ reservoir resides in the cold molecular layer just below the CO ice line with $\mathrm{\epsilon(N_2H^+)\gtrsim10^{-11}}$. The lower boundary of this layer with $\mathrm{\epsilon(N_2H^+)<10^{-11}}$ is reached at $T_\mathrm{d}\approx16\,\mathrm{K}$ where $\mathrm{\epsilon(N_2)\lesssim10^{-6}}$ due to freeze-out. Radially this layer extends from the inner midplane CO ice line out to $r\approx250-300\,\mathrm{au}$. Close to the midplane $\mathrm{\epsilon(N_2H^+)\lesssim10^{-12}}$ for $r\gtrsim150\,\mathrm{au}$ due to non-thermal desorption of ices. There is also a thin N$_2$H$^+$ layer at the top of the warm molecular layer with $\mathrm{\epsilon(N_2H^+)\approx10^{-12}}$ extending from the inner radius of the disk out to $r\approx100\,\mathrm{au}$. In this layer the \mbox{X-ray} ionization rate is high enough to compensate for the destruction of N$_2$H$^+$ by CO.

The detailed appearance of this layered structure is especially sensitive to the dust temperature and therefore also to dust properties (e.g. dust size distribution). The above described abundance structure for CO, HCO$^+$ and N$_2$H$^+$ is consistent with the model of \citet{Aikawa2015} that includes millimetre sized dust particles with a dust size distribution similar to what is used here (for details see Appendix~\ref{sec:chemistry}).  
\begin{figure*}
\resizebox{\textwidth}{!}{\includegraphics{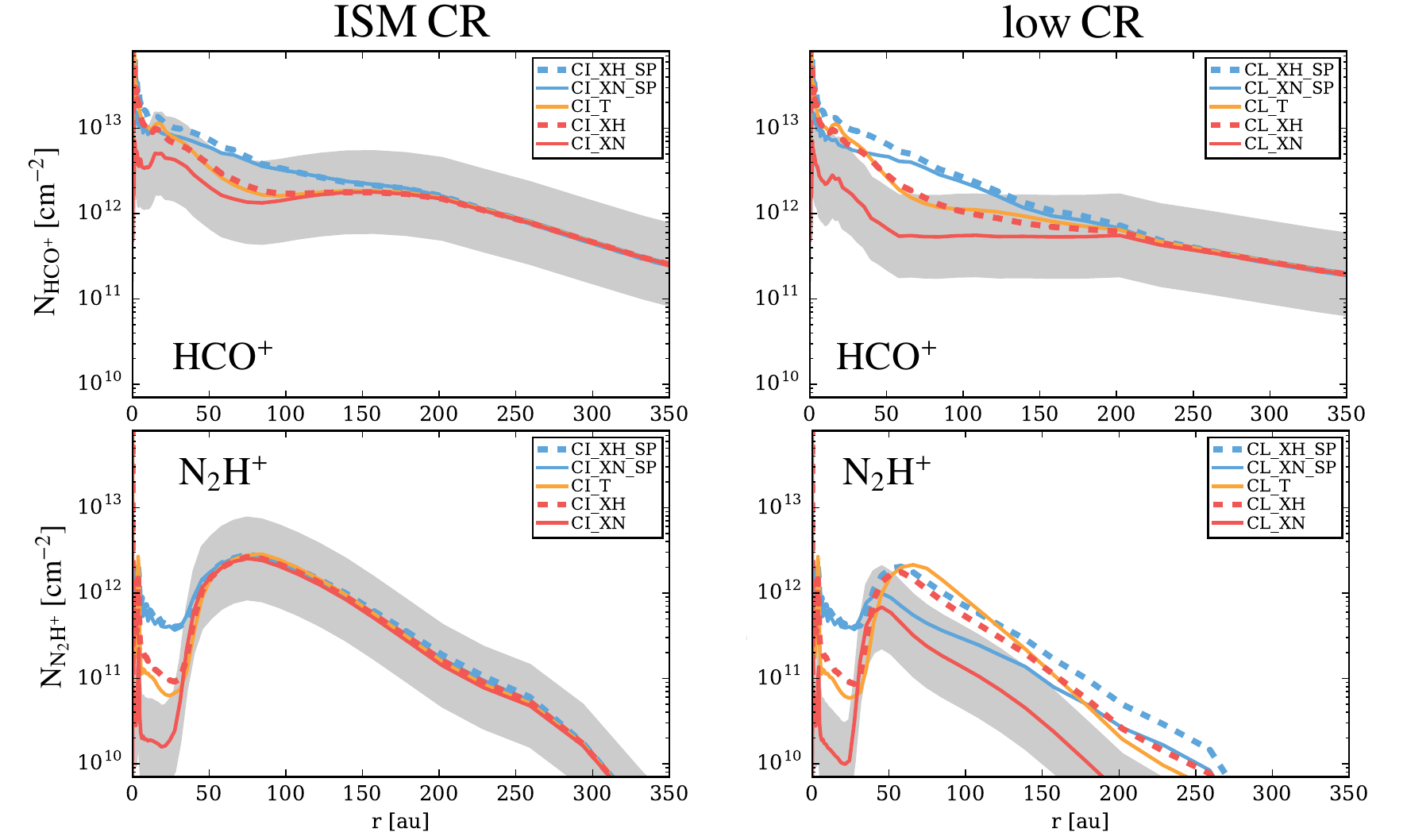}}
\caption{Vertical column density profiles for HCO$^+$ and N$_2$H$^+$ for our model series (Table~\ref{table:models}). The left column shows the models with the ISM like CRs ($\zeta_{\mathrm{CR}}\approx 2\times10^{-17}\,\mathrm{s^{-1}}$) the right column with low CRs ($\zeta_{\mathrm{CR}}\approx 2\times10^{-19}\,\mathrm{s^{-1}}$). The top row shows HCO$^+$, the bottom row N$_2$H$^+$. The blues lines are for models with, the red lines are for models without SPs. Dashed (solid) lines are for models with high (normal) X-rays. The orange solid line shows the Turner model. The gray shaded area marks a difference of a factor 3 in the column densities relative to the CI\_XN (ISM CR, normal X-rays) and CL\_XN model (low CR, normal X-rays), respectively.}
\label{fig:cd_comp}
\end{figure*}
\subsubsection{Vertical column densities}
\label{sec:vericalcd}
To study the impact of SP ionization quantitatively we compare vertical column densities of HCO$^+$ and N$_2$H$^+$ for models with and without SPs. In Fig.~\ref{fig:cd_comp} we show the vertical column densities $N\mathrm{_{HCO^+}}$ and $N\mathrm{_{N_2H^+}}$ as a function of the disk radius $r$ for all models listed in Table~\ref{table:models}. The left column in Fig.~\ref{fig:cd_comp} shows the models with ISM CRs, the right column the models with low CRs. At first we discuss the models without SPs and compare them to other theoretical models.

The $N_\mathrm{HCO^+}$ profile shows a dip around $r\approx50-100\,\mathrm{au}$ in the ISM CR models CI\_XN and CI\_XH (high X-rays). The dip is also seen in the models of \citet{Cleeves2014a}. Differently to \citet{Cleeves2014a}, in our model this dip is not predominantly due to the erosion of CO by reactions with He$^+$ \citep[``sink effect" e.g.][]{Aikawa1997b,Bergin2014,Furuya2014g,Aikawa2015}, but mainly due to the interplay of CO freeze-out and non-thermal desorption in the outer disk. The CO sink effect is also active in our model but less efficient (see Appendix~\ref{sec:N2EB}). 

The lack of HCO$^+$ in the disk midplane at $r\approx50-100\,\mathrm{au}$ due to CO freeze-out is also visible in the HCO$^+$ abundance structure shown in Fig.~\ref{fig:abun}. Non-thermal desorption in the midplane produces CO abundances $\gtrsim10^{-7}$ for $r \gtrsim150\,\mathrm{au}$ and consequently also a slight increase in $N_\mathrm{HCO^+}$. In the low CR models the ionization rate is too low to produce a significant amount of HCO$^+$ in the cold molecular layer and the dip in the profile vanishes.

N$_2$H$^+$ traces the distribution of gas phase CO as it is efficiently destroyed by CO \citep{Qi2013br,Qi2013bc,Qi2015y}. This is also seen in our model. The sharp transition in the N$_2$H$^+$ column density at $r\approx 30\,\mathrm{au}$ traces the onset of CO freeze-out (Fig.~\ref{fig:cd_comp}). We note however, that the actual midplane CO ice-line is at $\mathrm{\approx 12\,\mathrm{au}}$ (see Appendix~\ref{sec:timedepchem} for details). In the ISM CR models $N\mathrm{_{N_2H^+}}$ is dominated by CR ionization as N$_2$H$^+$ mainly resides in the cold molecular layer. In the warm N$_2$H$^+$ layer \mbox{X-ray}  ionization dominates. However, due to the lower densities in the warm molecular layer, this layer only contributes significantly to the column density within the radial CO ice line, and the impact of X-rays is only visible there (compare models CI\_XH and CI\_XN in Fig.~\ref{fig:cd_comp}). The high \mbox{X-ray} luminosity decreases the contrast between the peak of $N\mathrm{_{N_2H^+}}$ close to the radial CO ice line and $N\mathrm{_{N_2H^+}}$ inside the CO ice line by about a factor of five.

For the low CR case $N\mathrm{_{N_2H^+}}$ drops by more than an order of magnitude compared to the ISM CR case. Such a strong impact of CR ionization on $N\mathrm{_{N_2H^+}}$ is also reported by \citet{Aikawa2015} and \citet{Cleeves2014a}. Higher \mbox{X-ray} luminosities can compensate for low CR ionization only to some extent. In the high \mbox{X-ray} model CL\_XH, $N\mathrm{_{N_2H+}}$ is lower by a factor of five compared to the ISM CR models.
\begin{figure*}
\resizebox{\textwidth}{!}{\includegraphics{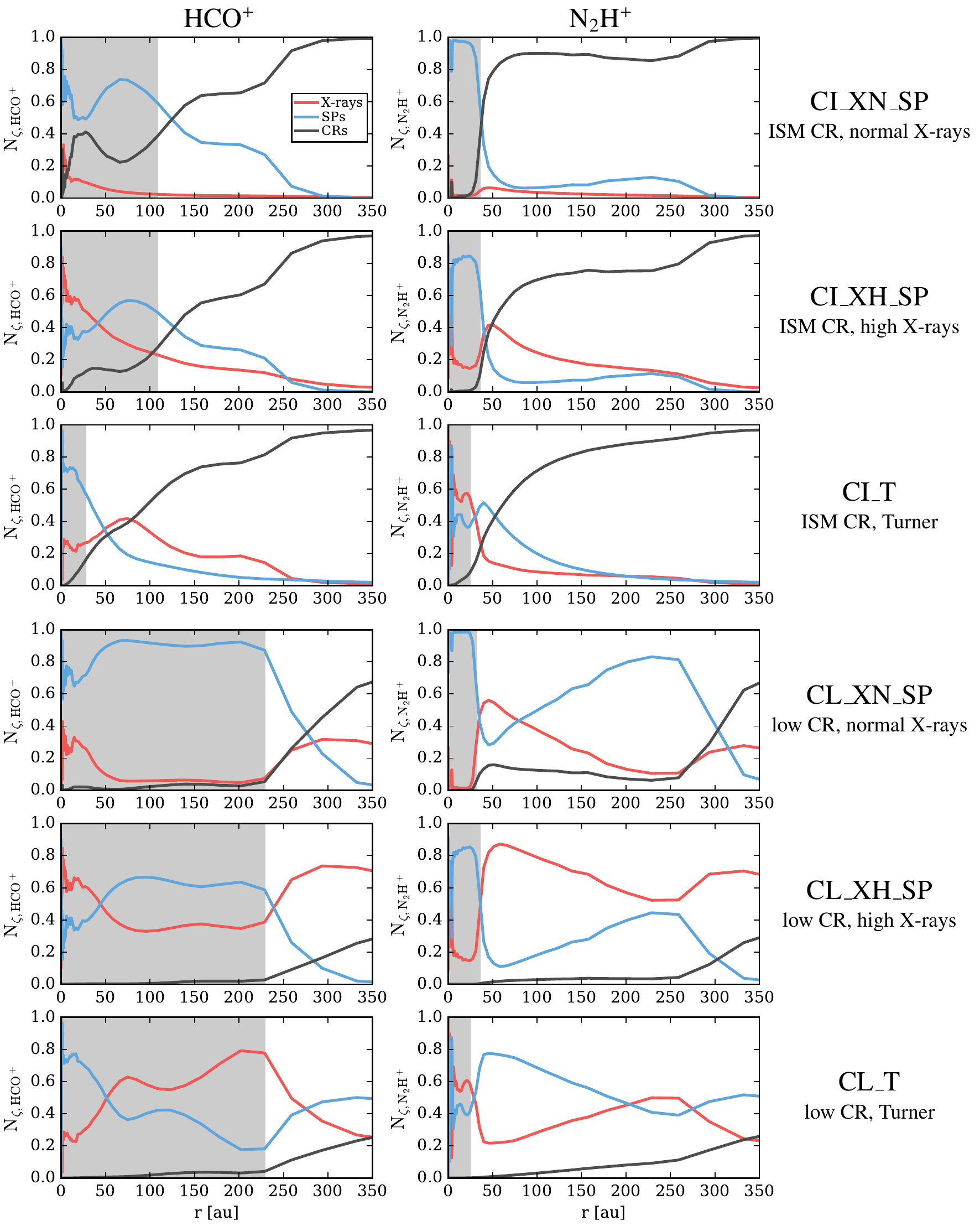}}
\caption{Ionization rate weighted column densities $\mathrm{N_\zeta}$ (Eq.~(\ref{eqn:nzeta})) as a function of radius for HCO$^+$ (left column) and N$_2$H$^+$ (right column). $\mathrm{N_\zeta}$ is normalized to the total column density of the respective molecule. The individual colored solid lines show the fraction of the total column density dominated by a certain ionization source. Red is for X-rays ($\mathrm{N_{\zeta_X}}$), blue for SPs ($\mathrm{N_{\zeta_{SP}}}$) and black for CRs ($\mathrm{N_{\zeta_{CR}}}$). Each row corresponds to one model. On the right hand side the model descriptions are provided (see Table~\ref{table:models}). The gray shaded area marks the region where more than 50\% of the column density arise from regions above the CO ice line (i.e. the warm molecular layer).}
\label{fig:cd_rel}
\end{figure*}
\subsubsection{Impact of SPs}
The column densities $N\mathrm{_{HCO^+}}$ and $N\mathrm{_{N_2H^+}}$ for models with SPs are shown in Fig.~\ref{fig:cd_comp} (blue solid and dashed lines). The solid orange line in Fig.~\ref{fig:cd_comp} shows the results for the Turner model. We discuss the Turner model separately in Sect.~\ref{sec:turnerimpact}. We define a change in the column densities by at least a factor of three compared to the reference model as significant. This is indicated by the gray area around the column density profiles of the reference models CI\_XN and CL\_XN.

To better quantify the impact of SPs compared to the competing H$_2$ ionization sources , X-rays and CRs, we introduce the weighted column density
\begin{equation}
\label{eqn:nzeta}
N_\mathrm{\zeta}(r)=\int^{\infty}_0 n(r,z)\times f_\zeta(r,z)\,\mathrm{d}z\,\,\mathrm{[cm^{-2}]}.
\end{equation}
$N_\zeta$ is the weighted column density for a particular H$_2$ ionization source $\zeta$, $n$ is the number density of a particular molecule in units of $\mathrm{cm^{-3}}$ and 
\begin{equation}
f_\zeta=\frac{\zeta}{\zeta_\mathrm{X}+\zeta_\mathrm{SP}+\zeta_\mathrm{CR}}.
\end{equation}
$N_\zeta$ represents the fraction of the column density dominated by a particular ionization source $\zeta$. 

In Fig.~\ref{fig:cd_rel} we show $N_\zeta$ for HCO$^+$ and N$_2$H$^+$ normalized to the total column density of the respective molecule as a function of radius. The gray shaded area in each plot roughly marks the region where more than 50\% of the total column density of the molecules arise from disk regions above the CO ice line (i.e. from the warm molecular layer).

SP ionization has a significant impact on the $N_\mathrm{HCO^+}$ profile in all our models. In the ISM CR model (CI\_XN\_SP) $N\mathrm{_{HCO^+}}$ increases by a factor $\approx3$ for $50\lesssim r\lesssim 100\,\mathrm{au}$ and the dip in the profile seen in the models without SPs (CI\_XN) vanishes (top left panel in Fig~\ref{fig:cd_comp}). In the low CR models this region increases to $25\lesssim r\lesssim150\,\mathrm{au}$ and $N\mathrm{_{HCO^+}}$ reaches values up to an order of magnitude higher compared to the CL\_XN model (top right panel in Fig~\ref{fig:cd_comp}). In the models with high X-rays, SP ionization still has an significant impact and $N\mathrm{_{HCO^+}}$ increases by up to a factor of three (dashed lines in Fig.~\ref{fig:cd_comp}).

This situation is also clearly visible in Fig.~\ref{fig:cd_rel} where we show $\mathrm{N_\zeta}$ as a function of radius (Eq.~\ref{eqn:nzeta}). Although X-rays can be the dominant ionization source close to the star, in all models $N\mathrm{_{HCO^+}}$ is dominated by SP ionization for $50\lesssim r \lesssim 100-200\,\mathrm{au}$. Fig.~\ref{fig:cd_rel} also shows that in the low CR models $N\mathrm{_{HCO^+}}$ is mainly built up in the warm molecular layer for $r\lesssim200\,\mathrm{au}$. In this region (gray area in Fig.~\ref{fig:cd_rel})  the warm molecular layer contributes more to the total column density than the cold molecular layer.

For N$_2$H$^+$ the picture is more complex. In the ISM CR models SPs have only very little impact on $N\mathrm{_{N_2H^+}}$. Only in the inner 30~au, within the radial CO ice line, the N$_2$H$^+$ profile is significantly affected. The reason for this is the high SP ionization rate $\zeta_\mathrm{SP} \gtrsim 10^{-12}\,\mathrm{s^{-1}}$ in the warm molecular layer of the disk close to the star. 
In this region the abundance ratio of HCO$^+$/N$_2$H$^+$ drops from $>10^3$, in the models without SPs, to around 10 to 100 in models with SPs. These high ratios can be explained by the efficient destruction of N$_2$H$^+$ by CO. However, due to the high $\zeta_\mathrm{SP}$ this destruction path becomes less important, and the molecular ion abundance are mainly determined by the balance between ionization and recombination. 

\citet{Ceccarelli2014d} reported a very low measured HCO$^+$/N$_2$H$^+$ ratio of $\approx3-4$ in the Class~0 source \mbox{\object{OMC-2 FIR 4}}. They explain this low ratio by the high ionization rates due to SPs ($\zeta_\mathrm{SP} > 10^{-14}$). In our disk model $\zeta_\mathrm{SP}$ in the  warm molecular layer is comparable, but the HCO$^+$/N$_2$H$^+$ ratio is $\gg4$. This higher ratio is due to the higher densities of $10^8 - 10^{10}\,\mathrm{cm^{-3}}$ and the stronger UV field in the warm molecular layer, compared to the physical conditions in \object{OMC-2 FIR 4}. This is in agreement with the chemical models presented in \citet{Ceccarelli2014d}. 

The impact of SPs on the N$_2$H$^+$ abundance in the warm layer can extend out to $r\approx200\,\mathrm{au}$ (similar to HCO$^+$). However, for $r>30\,\mathrm{au}$ this layer does not significantly contribute to the total N$_2$H$^+$ column density as $N\mathrm{_{N_2H^+}}$ is dominated by the high density layer below the vertical CO ice line. 

Beyond the radial CO ice line $N\mathrm{_{N_2H^+}}$ is dominated by CRs in the ISM CR models and is not affected by X-rays nor SPs (bottom left panel in Fig.~\ref{fig:cd_comp}). Compared to the ISM CR models, $N\mathrm{_{N_2H^+}}$ is lower  by a factor of a few around the peak and by more than an order of magnitude at larger radii in the low CR models. As a consequence the profile is also steeper. Although higher X-rays (CL\_XH model) and also SPs can to some extent compensate low CR ionization rates, the $N\mathrm{_{N_2H^+}}$ profile is still steeper and  $N\mathrm{_{N_2H^+}}$ is lower by a factor $\approx2-6$ compared to the ISM CR models for $r\gtrsim 30\,\mathrm{au}$.

From Fig.~\ref{fig:cd_rel} we see that only in the CL\_XN\_SP model SPs dominate $N\mathrm{_{N_2H^+}}$ for $r\gtrsim70\,\mathrm{au}$. Actually the importance of SPs increases with $r$ in this model. In the CL\_XN\_SP model SPs are the dominant ionization source in regions with \mbox{$N\mathrm{_{<H>,rad}}\lesssim 10^{25}\,\mathrm{cm^{-2}}$}. The $N\mathrm{_{<H>,rad}}=10^{25}\,\mathrm{cm^{-2}}$ iso-contour is below the vertical CO ice line for $r>70\,\mathrm{au}$ and the layer between the CO ice line and $N_\mathrm{<H>,rad}=10^{25}\,\mathrm{cm^{-2}}$ becomes thicker with radius (see Figures~\ref{fig:ionR_dom} and \ref{fig:abun}). This explains also the change in the slope of $N\mathrm{_{N_2H^+}}$ compared to models without SPs. 

In the high \mbox{X-ray} model, CL\_XH\_SP, the picture is quite different. $N\mathrm{_{N_2H^+}}$ is now dominated by \mbox{X-rays} for $r\gtrsim30\,\mathrm{au}$. \mbox{X-rays} are efficiently scattered towards the midplane and therefore $\zeta_\mathrm{X}>\zeta_\mathrm{SP}$ in the cold N$_2$H$^+$ layer (see also Fig.~\ref{fig:ionR_dom}). X-rays affect the cold N$_2$H$^+$ layer at all radii therefore the slope of $N\mathrm{_{N_2H^+}}$ is steeper compared to the model where SP dominate (compare the blue solid line with the red dashed line in Fig.~\ref{fig:cd_comp}).

Our results show that HCO$^+$ is always significantly affected by SP ionization but N$_2$H$^+$ only in models with low CRs and normal X-rays (for $r>30\,\mathrm{au}$). As SPs can only reach the upper layers of the cold molecular layer, N$_2$H$^+$ is less sensitive to SP ionization than HCO$^+$.

\subsubsection{Impact of SPs in the Turner model}
\label{sec:turnerimpact}
In the Turner model SP ionization is just a scaled up version of CR ionization where SPs can also penetrate the disk vertically (see Appendix~\ref{sec:turnermodel}). The results of the Turner models are also shown in Figures \ref{fig:cd_comp} and \ref{fig:cd_rel}.

In the ISM CR models there is no significant impact on HCO$^+$ and N$_2$H$^+$ by SP ionization. $\zeta_\mathrm{SP}$ is significantly lower at low column densities compared to our models (see Sect.~\ref{sec:resionrate}). Therefore X-rays are the dominant ionization source in the warm molecular layer and HCO$^+$ is not significantly affected by SP ionization. The slight increase in the HCO$^+$ column density is mainly due to the higher X-ray luminosity in the Turner model compared to our reference model with normal X-rays. Similar to our models CRs dominate in the cold molecular layer.

In the Turner model with low CRs, SPs become the dominant ionization source in the cold molecular layer as they also penetrate the disk vertically. In this layer $\zeta_\mathrm{SP}$ reaches values of $\approx10^{-17}\,\mathrm{s^{-1}}$ at $r\approx100\,\mathrm{au}$. However, also in the Turner model SPs cannot compensate for a low CR ionization rate as $\zeta_\mathrm{SP}\propto1/r^2$ (geometric dilution).  

The impact of SPs in the Turner model is rather limited and restricted to the cold molecular layer, in strong contrast to our models. The differences are mainly due to the assumptions concerning the SP transport. In the Turner model SPs hit the surface of the disk and penetrate the disk vertically, whereas we assume that SPs travel only along radial rays. However, both approaches are an approximation of a likely more complex picture of SP transport in disks. We discuss this in more detail in Sect.~\ref{sec:future}.
\section{Discussion}
\label{sec:discussion}
\subsection{Constraining the SP flux of T~Tauri stars}
SP ionization has a significant impact on the column densities of HCO$^+$ and N$_2$H$^+$ in all our models. To actually constrain the SP flux from observations it is necessary to disentangle the contribution of SP ionization from the competing ionization sources CRs and X-rays. 

From a chemical point of view all three ionization sources act the same way, they ionize molecular hydrogen and drive the molecular ion chemistry. However, they also show distinct differences in how they irradiate and penetrate the disk. CRs act like a background source and irradiate the disk isotropically, whereas X-rays and SPs originate from the star and act like a point source. A further difference is their energy distribution. Due to their high energies CRs and SPs tend to move on straight lines whereas (hard) X-rays also experience scattering during their interaction with the disk.  

Those differences in their irradiation properties and their energy distribution allow to disentangle their impact on the ion chemistry at different locations in the disk. Our models show that the stellar ionization sources are more effective closer to the star and at the surface layers of the disk (Fig.~\ref{fig:cd_rel}). X-rays can also become an important ionization source in the midplane of the disk but not for the whole disk as, roughly speaking, the ionization rate of stellar ionization sources is $\propto1/r^2$. CR ionization affects the whole disk, but is, in contrast to the stellar ionization sources, more important for the outer disk and the midplane/cold layers of the disk (this argument also holds for SLR ionization).

\begin{figure}
\centering
\resizebox{\hsize}{!}{\includegraphics{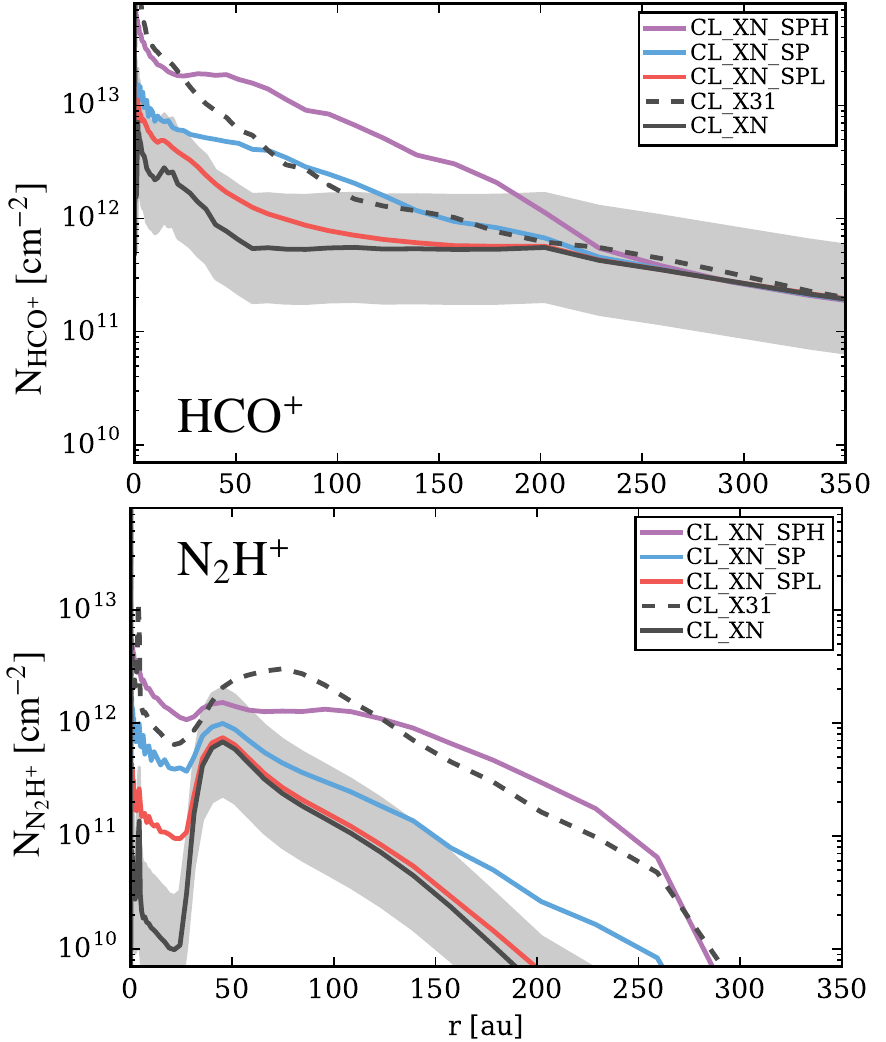}}
\caption{Comparison of column densities for HCO$^+$ (top) and N$_2$H$^+$ (bottom) for models with varying SP flux and very strong X-ray emission. For all models the low CR ionization rates are used. Shown are the reference model with normal X-rays (black, CL\_XN), the model with the typical SP flux (blue, CL\_XN\_SP) and models with a factor 10 higher (purple, CL\_XN\_SPH) and factor 10 lower (red, CL\_XN\_SPL) SP flux. The dashed black line shows the model with $L_\mathrm{X}=3\times10^{31}\,\mathrm{erg\,s^{-1}}$ and no SPs. The gray shaded area marks a difference of a factor 3 in the column densities relative to the reference model CL\_XN.}
\label{fig:cd_sphl}
\end{figure}
So far we only showed models using the commonly proposed SP flux of $f_\mathrm{p}(E_\mathrm{p}\ge 10~\mathrm{MeV})\approx10^7\,\mathrm{protons\,cm^{-2}\,s^{-1}}$  \citep[e.g.][]{Feigelson2002a}. However, this value should be only seen as an order of magnitude estimate (see Sect.~\ref{sec:sp}). 

In Fig.~\ref{fig:cd_sphl} we show the column densities for HCO$^+$ and N$_2$H$^+$ for models with a factor ten higher/lower SP flux with respect to the reference value. Also shown are the reference model for the low CR case CL\_XN and the model with the reference SP flux CL\_XN\_SP. For these models we use the low CR ionization rate ($\mathrm{\zeta_{CR}\approx2\times10^{-19}\,s^{-1}}$) and the normal X-ray luminosity ($L\mathrm{_{X}=10^{30}\,erg\,s^{-1}}$). Further we show a model with $L\mathrm{_X}=3\times10^{31}\,\mathrm{erg\,s^{-1}}$, to illustrated the impact of (very) strong X-ray emission (e.g. $\mathrm{\zeta_{X}\gtrsim10^{-17}\,s^{-1}}$ in the disk midplane).

In the low SP flux model (CL\_XN\_SPL, red line) the impact on HCO$^+$ and N$_2$H$^+$ is quite limited. For such a case it is still possible to define upper limits for the SP flux in disks. 

In the high SP flux model (CL\_XN\_SPH, purple line) the SP ionization rate reaches values of $\mathrm{\zeta_{SP}\gtrsim10^{-13}\,s^{-1}}$ and $\mathrm{\zeta_{SP}\approx10^{-17}\,s^{-1}}$ in the warm and cold molecular layer respectively. The $\mathrm{N_2H^+}$ column density profile beyond the radial CO ice line is comparable to the profile for models with ISM CRs (compare with Fig.~\ref{fig:cd_comp}). Also the model with very high X-rays (CL\_X31, black dashed line) and no SPs shows a similar profile for $\mathrm{N_2H^+}$. However, as seen from Fig.~\ref{fig:cd_sphl} the corresponding HCO$^+$ profiles differ significantly. 

Comparing the very high X-ray model (CL\_X31, black dashed line) to the reference model with SPs (CL\_XN\_SP, blue line) shows that the corresponding HCO$^+$ profiles are similar but the $\mathrm{N_2H^+}$ profiles differ by an order of magnitude. This shows again that it is indeed possible to distinguish between the different ionization sources by simultaneous modeling of HCO$^+$ and $\mathrm{N_2H^+}$ column density profiles. 

To trace this interplay of ionization sources, spatially resolved observations of molecular ion lines tracing different vertical layers of the disk are required. With modern (sub)millimetre interferometers like ALMA (Atacama Large Millimeter Array), NOEMA (NOrthern Extended Millimeter Array) and SMA (Submillimeter Array) such observations with a spatial resolution of tens of au are already possible \citep[e.g.][]{Qi2013bc,Cleeves2015a,ALMAPartnership2015,Yen2016} and will become available on a regular basis in the near future. Here, we use HCO$^+$ and N$_2$H$^+$ as the tracers of the warm and cold molecular layer respectively, but also other molecules like DCO$^+$, which traces similar layers/regions as N$_2$H$^+$ \citep{Teague2015b,Mathews2013v} can be used.

Complementary, far-infrared lines of HCO$^+$ and N$_2$H$^+$, as used by \citet{Ceccarelli2014d} to trace SP ionization in a protostellar envelope, are  good tracers of molecular ion emission in the warm inner region of the disk. However, a more detailed analysis with proper modeling of line emission is required to identify the best observational tracers of SP ionization. We will present such an analysis in a follow-up paper.
\subsection{Chemical implications}
Besides the H$_2$ ionization rates there are other ``chemical parameters'', which have an impact on the molecular ion abundance in disks. In the following we discuss the dependence of our results on the location of the CO ice line, depletion of CO and the assumed initial metal abundances. Those chemical properties of disks are not well constrained from observations and/or can vary between different targets. 
\subsubsection{Location of the CO ice line}
\label{sec:coiceline}
Recent ALMA observations provide direct constrains on the location of the CO ice line. However, these results depend on the method or more precisely the molecule used to trace the CO ice line (see \citealt{Qi2013bc,Schwarz2016,Nomura2016} for \object{TW Hya}). Further, due to complex chemical processes like the CO sink effect it is possible that the actual location of the CO ice line does not only depend on the CO freeze-out temperature (\citealt{Aikawa2015}, Sect.~\ref{sec:vericalcd}).

To investigate the dependence of our results on the location of the CO ice line we artificially move the CO ice line in our model by adapting the binding energy for CO. We consider two cases: $E\mathrm{_{B}(CO)=950\,K}$ and $E\mathrm{_{B}(CO)=1350\,K}$ (i.e $\mathrm{200\,K}$ lower and higher compared to our reference model). In both cases we keep the ratio of $\mathrm{E_B(N_2)/E_B(CO)}=0.67$ constant (see Appendix~\ref{sec:N2EB}). As a consequence also the $\mathrm{N_2H^+}$ layer moves accordingly to the CO ice line (see Sect.~\ref{sec:abunstruc}). For $E\mathrm{_{B}(CO)=950\,K}$ the CO ice line moves to $\mathrm{T_d\approx20-24\,K}$, (i.e. deeper into the disk) and for $E\mathrm{_{B}(CO)=1350\,K}$ to $T\mathrm{_d\approx25-36\,K}$ (i.e. higher up in the disk).

For a CO ice line deeper in the disk the contribution of CR ionization to the total column of HCO$^+$ and N$_2$H$^+$ increases in the ISM CR models. In the ISM CR models SP ionization is not significant anymore (i.e. $N\mathrm{_{HCO^+}}$ increases by less than a factor two). However, in the low CR models the impact of SPs remains significant.

A CO ice line higher up in the disk has the opposite effect. The total column densities of the molecular ions are now dominated by layers higher up in the disk which can efficiently be ionized by SPs. As a consequence the relative contribution of SP ionization to $N\mathrm{_{HCO^+}}$ and $N\mathrm{_{N_2H^+}}$ increases. 

In summary for a CO ice line location deeper in the disk SP ionization becomes less important, for a CO ice line higher up in the disk SP ionization becomes more important. However, in both cases the interplay of the different ionization sources is qualitatively speaking similar to what is shown in Fig.~\ref{fig:cd_rel}.
\subsubsection{CO depletion}
\label{sec:codepletion}
There is observational evidence for CO depletion in protoplanetary disks
\citep{Dutrey1997b,Bruderer2012b,Favre2013d,Kama2016,Schwarz2016,McClure2016}. The best constraint case is \object{TW Hya}. Using spatially resolved ALMA spectral line observations of several CO isotopologues \citet{Schwarz2016} derived a uniform CO abundance of $\approx10^{-6}$ in the warm molecular layer, two order of magnitudes lower than the canonical value of $\approx 10^{-4}$. However, the degree of CO depletion seems to vary from source to source. Using Herschel $\mathrm{HD}\,\mathrm{J}=1-0$ line observations \citet{McClure2016} derived CO depletions of a factor of $\approx 5$ and up to $\approx 100$ for \object{DM~Tau} and \object{GM~Aur}, respectively.

The cause of CO depletion in disks is not yet clear. Although freeze-out of CO certainly contributes to depletion it is unlikely that it is the only process acting. Several other mechanisms that can at least partly explain CO depletion are proposed:
\begin{itemize}
  \item the destruction of CO by He$^+$ and the subsequent conversion of atomic carbon to more complex carbon bearing molecules with higher freeze-out temperatures \citep{Aikawa1996h,Bergin2014,Helling2014,Furuya2014g}, 
  \item depletion of CO in layers above the CO ice line (up to $T\approx30\,\mathrm{K}$) due to conversion of CO to CO$_2$ on the surfaces of dust grains \citep{Reboussin2015b}, 
  \item  CO isotopologue selective photodissociation, which affects CO isotopologue line emission and therefore the derived CO depletion factors, 
  \item carbon and/or oxygen depletion in the warm disk atmosphere due to  settling and mixing of ice coated dust grains \citep{Du2015h,Kama2016a}.
\end{itemize}

It is yet unclear which of the proposed mechanisms is the most efficient one; none of them can be excluded with certainty. It is also possible that all of these processes are at work. So far the impact of CO (carbon/oxygen depletion) on molecular ion emission was not studied in detail. For the modeling of HCO$^+$ and N$_2$H$^+$ line emission of \object{TW Hya}, \citet{Cleeves2015a} reduced the initial atomic carbon abundance by two orders of magnitude to match C$^{18}$O line observation. However, the impact of C/CO depletion on HCO$^+$ and N$_2$H$^+$ was not discussed in detail.

To simulate CO depletion we simply reduce the total carbon and oxygen element abundances by one order of magnitude throughout the disk. This results in a CO abundance of $\approx 10^{-5}$ in the warm molecular layer. We applied this ``artificial'' CO depletion to all models listed in Table~\ref{table:models}; all other parameters of the models are fixed.

The CO depletion models show a factor of $\approx5-10$ lower $N\mathrm{_{HCO^+}}$ for $r\gtrsim50\,\mathrm{au}$ compared to the non depleted models. For $r<30\,\mathrm{au}$ $N\mathrm{_{N_2H^+}}$ increases by more than an order of magnitude. $N\mathrm{_{N_2H^+}}$  beyond the CO ice line is not affected as $N\mathrm{_{N_2H^+}}$ resides within the CO freeze-out zone where gas phase CO is anyway depleted.

In the CO depletion models SP ionization is slightly more efficient for $N\mathrm{_{N_2H^+}}$ as the contribution of the warm $\mathrm{N_2H^+}$ layer to $N\mathrm{_{N_2H^+}}$ increases. Due to the lower CO abundance in the warm molecular layer the $\mathrm{N_2H^+}$ abundance increases as the destruction pathway via CO is less efficient. For $\mathrm{HCO^+}$ the opposite is true. The $\mathrm{HCO^+}$ abundance decrease by roughly an order of magnitude in the warm molecular layer, whereas in the CO freeze-out the impact is smaller (i.e. CO is frozen-out anyway). Relatively speaking the contribution of the cold HCO$^+$ to $N\mathrm{_{HCO^+}}$ increases in the CO depletion models. Therefore the impact of SPs on $N\mathrm{_{HCO^+}}$ is less significant whereas X-rays and CRs become more important. Although there are some differences, the impact of SPs on $N\mathrm{_{N_2H^+}}$ and $N\mathrm{_{HCO^+}}$ is qualitatively very similar to the non-depleted models. In particular the main trends derived from Fig.~\ref{fig:cd_rel} are also seen in the CO depletion models.

CO depletion is certainly more complex than modelled here. A more thorough study of the impact of CO depletion on molecular ion abundances is certainly desirable and possibly provides new constraints on CO gas phase depletion in disks. However, this is beyond the scope of this paper.
\subsubsection{Metal abundances}
\label{sec:metalabun}
Heavy metals like sulphur play an important role in the molecular ion disk chemistry (\citealt{Teague2015b}, \mbox{Rab et al. in prep.}). Here we refer with the term metals to the elements \element{Na}, \element{Mg}, \element{Si}, \element{S} and \element{Fe}. Metal ionization due to UV radiation can produce a large number of free electrons. Those free electrons destroy molecular ions like HCO$^+$ and N$_2$H$^+$ via dissociative recombination. Dissociative recombination is more efficient than radiative recombination of metals. As a consequence a high abundance of metals significantly reduces the abundance of molecular ions \citep[e.g.][]{Mitchell1978a,Graedel1982c}. 

In the ISM and in disks most of the metals are likely locked up in refractory grains and are therefore depleted compared to Solar abundances. We use metal abundances similar to the commonly used ``low metal'' abundances \citep{Graedel1982c,Lee1998}. These low metal abundances are depleted by a factor $\approx100-1000$ compared to Solar abundances (i.e. low metal sulphur abundance $\mathrm{\epsilon(S)\approx10^{-7}}$). However, the actual gas phase abundance of metals is difficult to constrain from observations and a stronger degree of depletion already prior to disk formation is possible \citep[e.g.][]{Maret2007a,Maret2013a}.

To investigate the dependence of our results on the metal abundances we deplete the initial gas phase metal abundance by an additional factor of 10 compared to the low metal abundances (i.e. $\mathrm{\epsilon(S)\approx10^{-8}}$). This means that a larger fraction of metals is locked-up in refractory dust grains and cannot be released back into the gas phase.

Decreasing the metal abundances increases $\mathrm{\epsilon(HCO^+)}$ in the warm molecular layer by up to an order of magnitude (i.e. the gap in the vertical $\mathrm{\epsilon(HCO^+)}$ profile nearly vanishes; Sect.~\ref{sec:abunstruc}). In the cold molecular layer the metal abundances are not as important since the metals are frozen-out on dust grains anyway. However, in regions where non-thermal desorption processes are efficient (i.e. metal ices are released back into the gas-phase), $\mathrm{\epsilon(HCO^+})$ but also $\mathrm{\epsilon(N_2H^+)}$ are higher by a factor of a few in the lower metal abundance models. The HCO$^+$ column density is higher by a factor of $\approx2-3$ at all radii in the strong metal depletion models compared to the models with the reference  abundances. The N$_2$H$^+$ column density is only affected for $r\gtrsim150\,\mathrm{au}$ (higher by a factor of $\approx2-3$), where non-thermal desorption of metals is efficient. 

Lower gas phase metal abundances lead to an increase of the molecular ion abundances in the warm molecular layer and the contribution of this layer to the total column densities increases. However, in the the strong metal depletion  models the interplay and the relative contributions to the column densities of the different ionization sources is nearly identical to the reference model grid (i.e. Fig.~\ref{fig:cd_rel} does not change significantly). Our arguments concerning the impact of SPs on HCO$^+$ and N$_2$H$^+$ are therefore also valid for the case of strong metal depletion.
\subsection{Future prospects}
\label{sec:future}
Our model results show that SPs can indeed become an important ionization source in T~Tauri disks. However, our model should be seen as a first approach towards a comprehensive modeling of SP ionization in protoplanetary disk. In the following we discuss further important aspects like variability, non-stellar origin of SPs, the importance of magnetic fields and future prospects for SP modeling in protoplanetary disks.
\subsubsection{Flares and variability}
In our models we assumed continuous (i.e. time-averaged) particle fluxes and X-ray luminosities. Although this is a reasonable assumption (see Sect.~\ref{sec:sp}) it is likely that the disk is also hit by singular powerful X-ray and/or SP flares. 

From the X-ray COUP survey of the ONC cloud, \citet{Feigelson2002a} and \citet{Wolk2005b} derived a median X-ray flare luminosity of $\approx6\times10^{30}\,\mathrm{ergs\,s^{-1}}$, comparable to the X-ray luminosity in our high X-ray models, and peak flare luminosities up to $10^{31}\,\mathrm{ergs\,s^{-1}}$ . The duration of such flares can last from hours up to three days with a typical frequency of roughly one powerful flare per week. 

\citet{Ilgner2006f} argue that for such flare properties the disk ion chemistry responds to time-averaged ionization rates. However, in case the duration between two flares is longer than the recombination time scale in the disk, such singular and strong flares would produce asymmetric features in molecular ion emission (i.e. a singular flare only affects a certain fraction of the disk). The spatially resolved $\mathrm{HCO^+\,J=3-2}$ SMA observations of \object{TW Hya} show indeed such an asymmetric structure \citep{Cleeves2015a}. However, as discussed by \citet{Cleeves2015a} these features could also have a different origin like spiral arms or an hidden planet locally heating the disc.

If such asymmetric structures are caused by stellar flares they would provide complementary constraints on the X-ray/SP activity of the star. Multi-epoch data of spatially resolved molecular ion emission is required to prove the flare scenario (i.e. the features should disappear quickly). The current (sub)mm interferometers like ALMA, NOEMA and SMA provide the required spatial resolution and might even allow for monitoring of disks in molecular ion lines on a daily/weekly basis in the future.

Modelling of such observations does not necessarily require 3D chemical models. Assuming that the disk physical structure is not affected by flares, radial cuts through the disk can be modelled with 2D (time-dependent) thermo-chemical models as presented here. 
\subsubsection{Magnetic fields}
\label{sec:mangeticfields}
In our model we neglect the impact of magnetic fields on the SP transport. Stellar and disk magnetic fields are in particular relevant for the question if particles actually hit the disk (see \citealt{Feigelson2002a} for a discussion). 

Magnetic fields can either drag the particles away from the disk (e.g. like in a wind) but could also funnel the particles and concentrate the ionizing flux in particular regions of the disk. In the first scenario SPs can still have an impact on the upper layers of the disk but certainly not on the disk midplane. In the second scenario particles are likely focused on regions close to the inner radius of the disk and their impact on the outer disk will become less significant. Also the trajectory of the particles will be affected, and they might penetrate the disk also vertically. Our models, where particles are transported only radially, are closer to the wind scenario, whereas the Turner model would represent an extreme case for magnetically focused particles. However, to qualitatively estimate the impact of magnetic fields more complex SP transport models are required.

It is possible to consider magnetic field effects in high energy particle transport models \citep{Desch2004,Padovani2011c,Padovani2013}. In principle such methods can also be applied to disk models. The main challenge though is to determine the structure of the star and disk magnetic fields. However, as argued by \citet{Ceccarelli2014d} identifying distinct observational signatures of SP ionization in disks and/or envelopes would allow to derive constrains for the magnetic field structure. This is certainly challenging, but with the availability of spatially resolved observational data and interpretation of such data with (improved) thermo-chemical disk models this might be feasible in the near future.
\subsubsection{Non-stellar origin of high energy particles}
Besides the stellar surface the close environment of young stars offers also alternative particle acceleration sites. X-ray flares and particles can be produced close to the inner disk in the so called reconnection ring where the stellar and disk magnetic field interact \citep[X-wind model][]{Shu1994,Shu1997q}. More recently \citet{Padovani2015f,Padovani2016b} proposed jet shocks as alternative acceleration sites for high energy particles. 

In the X-wind model the particle source and also the X-ray emitting source is located closer to the disk and slightly above the disk midplane. The typically assumed source location in this ``lamppost'' scenario is $r_\mathrm{L}\approx0.05\,\mathrm{au}$ and $z_\mathrm{L}\approx0.05\,\mathrm{au}$ ($\approx10\,\mathrm{R_\sun}$) \citep{Lee1998mq,Igea1999}. For X-rays \citet{Ercolano2009b} found that the height of the emitting source has relatively little impact concerning X-ray radiative transfer in disks. Moving the emitting source closer to the disk certainly has an impact on the very inner disk but for e.g. $10\,\mathrm{au}$ distance from the star the stellar particle flux increases only by about 1\% compared to the stellar origin. The height of the SP emitting source has some impact on where in the disk SPs see a column density $N\mathrm{_{<H>}}\gtrsim10^{25}\,\mathrm{cm^{-2}}$. For particles accelerated at a height $z_\mathrm{L}\approx0.05\,\mathrm{au}$ and moving along a ray parallel to the midplane this happens at $r\approx 0.5\,\mathrm{au}$. The consequences are that SPs can penetrate into slightly deeper layers of the disks but they still cannot penetrate to the disk midplane. Therefore our conclusions on the impact of SPs on HCO$^+$ and N$_2$H$^+$ remain valid for the X-wind scenario. 

In the jet shock scenario, the emitting source would be located far above the disk. \citet{Padovani2016b} considered a particle emitting source located at $1.8\times10^3\,\mathrm{au}$ above the star and calculated the resulting ionization rates for a 2D disk structure. Depending on their particle acceleration model they found ionization rates up to $\zeta\approx10^{-14}\,\mathrm{s^{-1}}$ at the surface layers of the disk. Although X-ray ionization rates in the upper layer of the disk are typically higher such an irradiation scenario could have a significant impact on the ionization of the outer disk ($r\gtrsim 50\,\mathrm{au}$). In the outer disk, the jet accelerated particles can penetrate the disk vertically and therefore can also reach the disk midplane (similar to the Turner model, Sect.~\ref{sec:turnerimpact}). This might become important if Galactic CRs are efficiently attenuated (i.e. in the low CR case).

The scenarios described above for non-stellar particle sources are certainly worth being investigated in detail. For the future we plan to extend our model to a proper treatment of such non-stellar emitting sources.
\section{Summary and conclusions}
\label{sec:conclusions}
In this work we investigated the impact of stellar energetic particle (SP) ionization on disk chemistry with a focus on the common disk ionization tracers HCO$^+$ and N$_2$H$^+$. We assumed a typical SP flux of $f_\mathrm{p}(E_\mathrm{p}\approx 10~\mathrm{MeV})\approx10^7\,\mathrm{protons\,cm^{-2}\,s^{-1}}$ (at $1\,\mathrm{au}$) as commonly proposed for T~Tauri stars and a particle energy distribution derived from measurements of solar particle events. Based on a detailed particle transport model we derived an easy to use formula (see Sect.~\ref{sec:sp}) to calculate the SP ionization rate in the disk as a function of hydrogen column density and radius, assuming that the particles can penetrate the disk only radially. With a small grid of models considering varying properties of the competing high energy disk ionization sources, X-rays and Galactic cosmic rays, we studied the interplay of the different ionization sources and identified possible observational tracers of SP ionization.
Our main conclusions are the following: 
\begin{itemize}
  \item SPs cannot penetrate the disk midplane. At hydrogen column densities $N\mathrm{_{<H>}\gtrsim10^{25}\,cm^{-2}}$ even the most energetic particles are attenuated (stopped) and the SP ionization rate drops rapidly. As the radial hydrogen column densities for full T~Tauri disks are typically $N\mathrm{_{<H>}\gg10^{25}\,cm^{-2}}$ the midplane SP ionization rate is $\mathrm{\zeta_{SP}\ll10^{-20}\,s^{-1}}$ already at distance of $1\,\mathrm{au}$ from the star. 
  \item For the assumed SP flux (see above), SPs become the dominant H$_2$ ionization source in the warm molecular layer of the disk above the CO ice line, provided that SPs are not shielded by magnetic fields. This is even true for enhanced X-ray luminosities (i.e. $L\mathrm{_X}=5\times10^{30}\,\mathrm{erg\,s^{-1}}$).
  \item SP ionization can increase the HCO$^+$ and N$_2$H$^+$ column densities by factors of $\approx3-10$ for disk radii $r\lesssim200\,\mathrm{au}$. The impact is more significant in models with low CR ionization rates (i.e. $\mathrm{\zeta_{CR}\approx10^{-19}\,s^{-1}}$). 
  \item SP ionization becomes insignificant for an SP flux one order of magnitude lower than the proposed value for T~Tauri stars. In such a case H$_2$ ionization is solely dominated by X-rays and CRs.  
  \item As SPs cannot penetrate the deep layers of the disk, \mbox{X-rays} and/or CRs usually remain the dominant H$_2$ ionization source in the cold disk layers (i.e. below the CO ice line). Therefore HCO$^+$, which traces the warm molecular layer, is more sensitive to SP ionization than N$_2$H$^+$ that resides in the cold molecular layer.
  \item Simultaneous modeling of spatially resolved radial intensity profiles of molecular ions tracing different vertical layers of the disk allows to disentangle the contributions of the competing high energy ionization sources to the total H$_2$ ionization rate. Consequently such observations allow to constrain the SP flux in disks. Such a method is likely model dependent and ancillary observations constraining the vertical chemical structure of disks are required.
\end{itemize}
We have shown that stellar energetic particles can be an important ionization agent for disk chemistry. Modelling of spatially resolved observations of molecular ions with a model such as presented here allows to put first constraints on the stellar particle flux in disks around T~Tauri stars. 

Further model improvements concerning the stellar energetic particle transport (i.e. magnetic fields) are required to answer the question to what extent stellar particles reach the disk. Additionally non-stellar origins (i.e. jets) of high energy particles should be considered. With such models and spatially resolved molecular ion observations it will be possible to put stringent constraints on stellar energetic particle fluxes of T~Tauri stars and to infer properties of the stellar and disk magnetic fields.

\begin{acknowledgements} The authors thank the anonymous referee for useful suggestions and comments. The research leading to these results has received funding from the European Union Seventh Framework Programme FP7-2011 under grant agreement no 284405. RCH acknowledges funding by the Austrian Science Fund (FWF): project number P24790. MP acknowledges funding from the European Unions Horizon 2020 research and innovation programme under the Marie Sk\l{}odowska-Curie grant agreement No 664931. The computational results presented have been achieved using the Vienna Scientific Cluster (VSC). This publication was supported by the Austrian Science Fund (FWF).
\end{acknowledgements}
\bibliographystyle{aa}
\bibliography{stcr}
\appendix
\section{X-ray radiative transfer}
\label{sec:app_xrt}
\begin{figure}
\resizebox{\hsize}{!}{\includegraphics{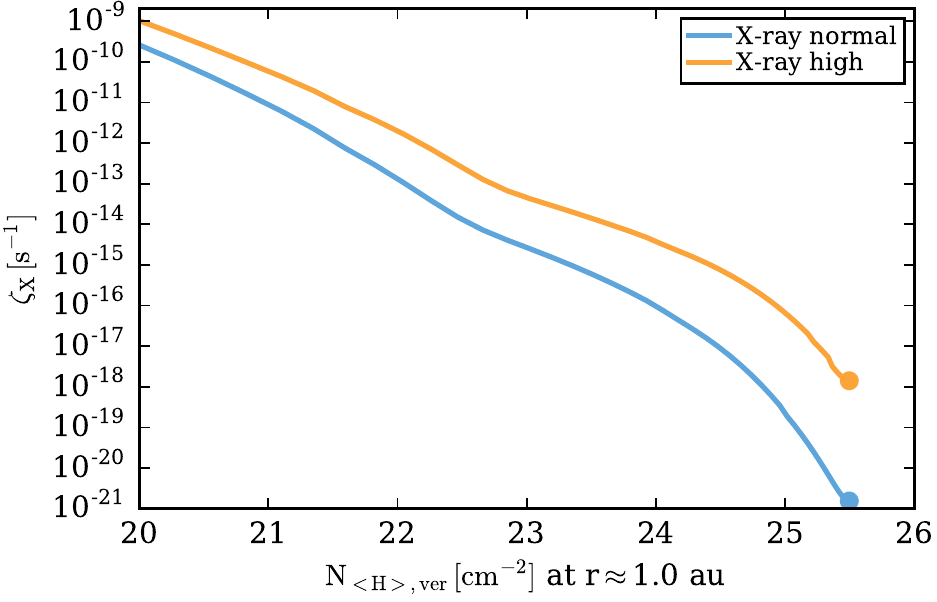}}
\caption{X-ray ionization rate $\zeta_\mathrm{X}$ as a function of vertical hydrogen column density of the disk at a distance of 1 au from the star. Shown are the resulting ionization rates for the two different X-ray spectra (see Sect.~\ref{sec:xrays}). The dots mark the same location ($r=1\,\mathrm{au}$, $z=0\,\mathrm{au}$) in the disk as the dots in Fig.~\ref{fig:xrayion_midplane}.}
\label{fig:xrayion_1au}
\end{figure}
We use here the newly developed X-ray radiative transfer module of P{\tiny RO}D{\tiny I}M{\tiny O}. The details of this module will be presented in a forthcoming paper \mbox{Rab et al. (in prep.)}. Here we only briefly describe the implementation and present results concerning the X-ray ionization rate for comparison with other models.

The main difference of the new model to the existing implementation of \citet{Aresu2011,Meijerink2012a} is the treatment of X-ray scattering. We use the absorption and scattering cross-section from the publicly available \emph{xraylib} library (\citealt{Schoonjans2011a}, \url{https://github.com/tschoonj/xraylib}). For the X-ray radiative transfer we use the same numerical approach (e.g. use of wavelength bands) as is used for the dust radiative transfer in P{\tiny RO}D{\tiny I}M{\tiny O} \citep{Woitke2009a}. We assume isotropic scattering but apply a correction factor (anisotropic factor g) to account for anisotropic Compton scattering. According to \citet{Cleeves2016a}, neglecting anisotropic scattering has only a limited impact on the X-ray ionization rate of about a factor of 2. We consider this possible deviation as not significant for the results presented here. We further neglect the dust in the X-ray radiative transfer, since with the assumed gas to dust ratio of 100 and dust settling X-ray photons mainly interact with the gas \citep{Bethell2011a,Glassgold2012}.

The X-ray chemistry in P{\tiny RO}D{\tiny I}M{\tiny O} is presented in detail in \citet{Meijerink2012a}. The interaction of X-rays with the gas changes the species abundances and consequently also the \mbox{X-ray} opacities. We therefore iterate between the X-ray radiative transfer and the chemistry until convergence is reached.

\begin{figure}
\resizebox{\hsize}{!}{\includegraphics{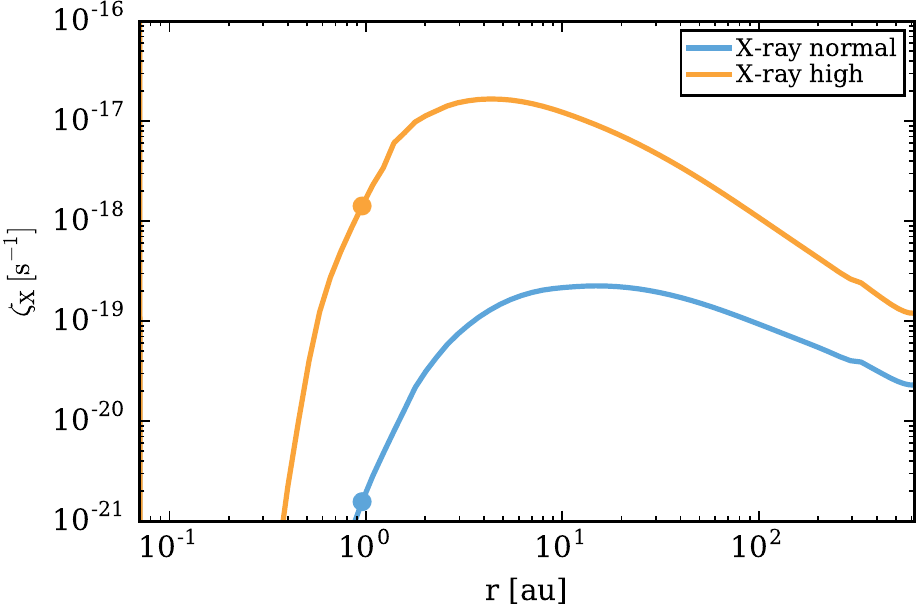}}
\caption{X-ray ionization rate $\zeta_\mathrm{X}$ as a function of radius in the midplane of the disk (z=0). Shown are the resulting ionization rates for the two different X-ray spectra (see Sect.~\ref{sec:xrays}). The dots mark the same location ($r=1\,\mathrm{au}$, $z=0\,\mathrm{au}$) in the disk as the dots in Fig.~\ref{fig:xrayion_1au}.}
\label{fig:xrayion_midplane}
\end{figure}

In Fig.~\ref{fig:xrayion_1au} and Fig.~\ref{fig:xrayion_midplane} we show the X-ray ionization rate $\zeta_\mathrm{X}$ for our disk model (Sect.~\ref{sec:diskmodel}) and for both X-ray spectra (Sect.~\ref{sec:xrays}). Fig.~\ref{fig:xrayion_1au} shows $\zeta_\mathrm{X}$ as a function of vertical column density at 1~au whereas Fig.~\ref{fig:xrayion_midplane} shows $\zeta_\mathrm{X}$ in the midplane of the disk as a function of radius. These results are in good quantitative agreement with \citet{Ercolano2013b} and \citet{Cleeves2015a} considering that  different disk models and different implementations of X-ray radiative transfer (e.g. cross-sections) are applied.
\section{Chemistry}
\label{sec:chemistry}
\begin{table*}
\caption{Rate coefficients for the main formation and destruction pathways of HCO$^+$ and N$_2$H$^+$.}
 \label{table:ratecoeff}
 \centering
 \begin{tabular}{lcccccc}
 \hline\hline
 Reaction & $\mathrm{\alpha}$ & $\mathrm{\beta}$ & $\mathrm{\gamma}$ & \tablefootmark{a}$k\mathrm{(25 K)}$ & Type & Reference \\
 & & & & $\mathrm{(cm^3\,s^{-1})}$ & & \\
 \hline
 $\mathrm{H_3^+ + CO \rightarrow H_2 + HCO^+}$ & $1.35(-9)$ & $-0.14$ & $-3.4$ & $2.2(-9)$ & Ion-neutral& 1 \\
 $\mathrm{HCO^+ + e^- \rightarrow H + CO}$ & $2.40(-7)$ & $-0.69$ & 0.0& $1.3(-6)$& Dissociative recombination & 2 \\
 $\mathrm{HCO^+ + H_2O \rightarrow CO + H_3O^+}$ & $2.50(-9)$& $-0.50$ & 0.0& $8.7(-9)$ & Ion-neutral & 3 \\
 $\mathrm{H_3^+ + N_2 \rightarrow H_2 + N_2H^+}$ & $1.80(-9)$ & 0.0& 0.0& $1.8(-9)$ & Ion-neutral & 4 \\
 $\mathrm{N_2H^+ + e^- \rightarrow H+ N_2}$ & $2.77(-7)$& $-0.87$& 0.0& $2.4(-6)$& Dissociative recombination & 5 \\
 $\mathrm{N_2H^+ + e^- \rightarrow NH+ N}$ & $2.09(-8)$& $-0.74$ & 0.0& $1.3(-7)$& Dissociative recombination & 5\\
 $\mathrm{N_2H^+ + CO \rightarrow N_2 + HCO^+}$ & $8.8(-10)$& 0.0& 0.0& $8.8(-10)$& Ion-neutral & 6 \\
 \hline
 \end{tabular}
 \tablefoot{The rate coefficient is given by the modified Arrhenius equation $k(T)=\alpha\times(T/300\,\mathrm{K})\times\exp(-\gamma/T)$ \\
 \textbf{References.} UMIST 2012 database \citep{McElroy2013b}; (1) \citet{Klippenstein2010}; (2) \citet{Mitchell1990}; (3) \citet{Adams1978}; (4) \citet{Rakshit1982}; (5) \citet{Lawson2011}; (6) \citet{Payzant1975,Herbst1975,Bohme1980}
 }
\end{table*}
For the chemical reaction network we use the UMIST~2012 database for gas phase chemistry \citep{McElroy2013b} for a selection of 235 species. Additionally the network includes X-ray \citep{Aresu2011,Meijerink2012a} chemistry, PAH (polycyclic aromatic hydrocarbons) chemistry \citep{Thi2014}, adsorption and thermal/non-thermal desorption (CR and photo desorption) for ice species and H$_2$ formation on grains \citep{Woitke2009a}. In total the network includes 3143 chemical reactions. For the ice species we use the binding energies from the UMIST~2012 release but updated a couple of values for oxygen bearing species (see Sect.~\ref{sec:EB}). Further details on the network can be found in \mbox{Kamp et al. (in prep.)}. The most relevant gas phase reactions for HCO$^+$ and N$_2$H$^+$ with their rate coefficients are listed in Table~\ref{table:ratecoeff}.

\begin{figure}
\centering
\resizebox{\hsize}{!}{\includegraphics{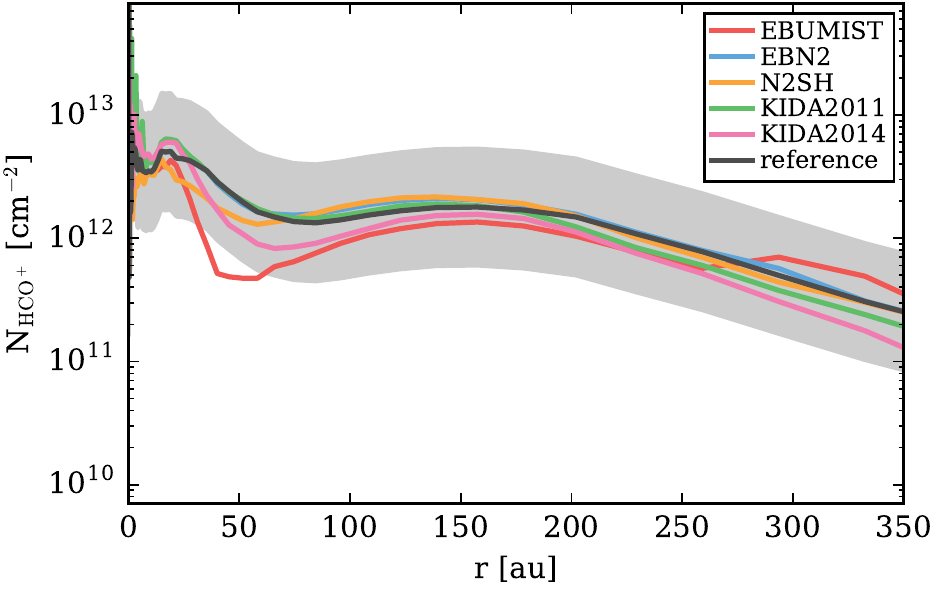}}\\
\vspace{-0.36cm}
\resizebox{\hsize}{!}{\includegraphics{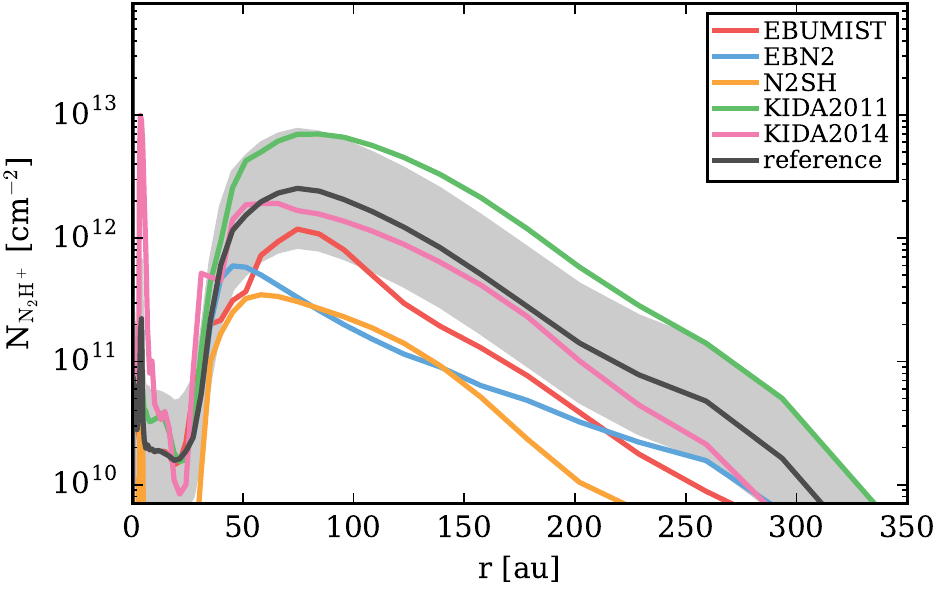}}
\caption{Comparison of column densities for HCO$^+$ and N$_2$H$^+$ for models with different chemical networks or binding energies (see Table~\ref{table:chemmodels}). The gray shaded area marks a difference of a factor 3 in the column densities relative to the reference model (black line).}
\label{fig:chemtest}
\end{figure}

To test the robustness of our chemical model with respect to HCO$^+$ and N$_2$H$^+$ we performed chemical tests with time-dependent chemistry, varying binding energies and the KIDA chemical network (Kinetic Database for Astrochemistry, \citealt{Wakelam2012a,Wakelam2015}). The different test models are described in Table~\ref{table:chemmodels} and discussed in the following sections. The results for the HCO$^+$ and N$_2$H$^+$ column densities are shown in Fig.~\ref{fig:chemtest} and Fig.~\ref{fig:chemtest_td}.
\subsection{N$_2$ shielding}
Dust shielding of N$_2$ photodissociation is strongly reduced in protoplanetary disks due to dust growth/evolution and shielding by H$_2$ becomes important \citep{Li2013}. To account for this we implemented the H, H$_2$ shielding and self-shielding functions for N$_2$ of \citep{Li2013} in our chemistry model. The data are taken from the Leiden photodissociation database \url{http://home.strw.leidenuniv.nl/~heays/photo/}. 

We find that N$_2$ shielding is very important for the abundance structure of N$_2$H$^+$. In models without N$_2$ shielding the vertical column density of N$_2$H$^+$ is reduced by one order of magnitude throughout the disk. Further the warm N$_2$H$^+$ layer above the vertical CO ice-line vanishes.      
\begin{table}
\caption{Chemical test models.}
\label{table:chemmodels}
\centering
\begin{tabular}{ll}
\hline\hline
Name & Description \\
\hline
EBUMIST      & original binding energies from \\
             & UMIST~2012 \\
EBN2         & N$_2$ to CO binding energy ratio  \\
             & of 0.9 according to experiments \\
N2SH         & no N$_2$ shielding \\
KIDA2011     & gas phase chemical network from \\ 
             & the KIDA 2011 release \\
KIDA2014     & gas phase chemical network from \\ 
             & the KIDA 2014 release \\
\hline
\end{tabular}
\tablefoot{All chemical test models are based on the reference model CI\_XN.}
\end{table}
\begin{table}
\caption{Important binding energies.}
\center
\label{table:adsenergies}
\begin{tabular}{lcc}
\hline \hline
Species & $E_\mathrm{B}$ UMIST 2012 & $E_\mathrm{B}$ \\
 & (K) & (K) \\ 
\hline
CO       & 1150 \tablefootmark{a} & \\    
N$_2$    & 790  \tablefootmark{b} & \\               
O        & 800  \tablefootmark{c} & 1500 \tablefootmark{e} \\
O$_2$    & 1000 \tablefootmark{d} & 1250 \tablefootmark{f} \\
O$_3$    & 1800 \tablefootmark{d} & 2100 \tablefootmark{g} \\
OH       & 2850 \tablefootmark{d} & 4600 \tablefootmark{h} \\
\hline
\end{tabular}
\tablefoot{
The second column shows the values from the UMIST 2012 database \citep{McElroy2013b}, the third column the updated values from \citet{Minissale2016a}. \newline \textbf{References.}  
\tablefoottext{a}{\citet{Garrod2006b,Collings2004d}}
\tablefoottext{b}{\citet{Oberg2005}}
\tablefoottext{c}{\citet{Tielens1987}}
\tablefoottext{d}{\citet{Garrod2006b}}
\tablefoottext{e}{\citet{Bergeron2008}}
\tablefoottext{f}{\citet{Noble2012b}}
\tablefoottext{g}{\citet{Borget2001,Minissale2014a}}
\tablefoottext{h}{\citet{Dulieu2013}}
}
\end{table}
\subsection{Binding energies}
\label{sec:EB}
\subsubsection{Oxygen binding energy}
Several laboratory experiments \citep[e.g.][]{Ward2012,He2014,Minissale2016b,Minissale2016a} reported binding energies for oxygen in the range of $E\mathrm{_B(O)}=1500-1800\,\mathrm{K}$. This is significantly higher than the value of $E\mathrm{_B(O)}=800\,\mathrm{K}$ listed in the UMIST~2012 database. We therefore updated the binding energies for oxygen and several other oxygen bearing species with the values listed in \citet{Minissale2016a}. The new values with their references are given in Table~\ref{table:adsenergies}.

The higher binding energy for oxygen has a significant impact on 
N$_2$H$^+$. Using the UMIST~2012 binding energy reduces the N$_2$H$^+$ column density by about a factor of three for $r\gtrsim 50\,\mathrm{au}$ (see model EBUMIST in Fig.~\ref{fig:chemtest}). Similar to CO, oxygen also destroys N$_2$H$^+$ via the reaction $\mathrm{O+N_2H^+\rightarrow N_2+ OH^+}$. Due to this reaction the N$_2$H$^+$ abundance is strongly reduced in a thin layer \emph{below} the vertical CO ice line and \emph{above} the vertical oxygen ice line. With the higher binding energy for oxygen this thin layer vanishes as oxygen freezes-out at higher temperatures than CO.

For HCO$^+$ only the dip in the column density profile at $r\approx50\,\mathrm{au}$ is significantly affected. In this region HCO$^+$ is efficiently destroyed by water. Removing oxygen and OH from the gas phase reduces also the water abundance near the vertical CO ice line, consequently the HCO$^+$ abundance increases.            
\begin{figure*}
\resizebox{0.48\hsize}{!}{\includegraphics{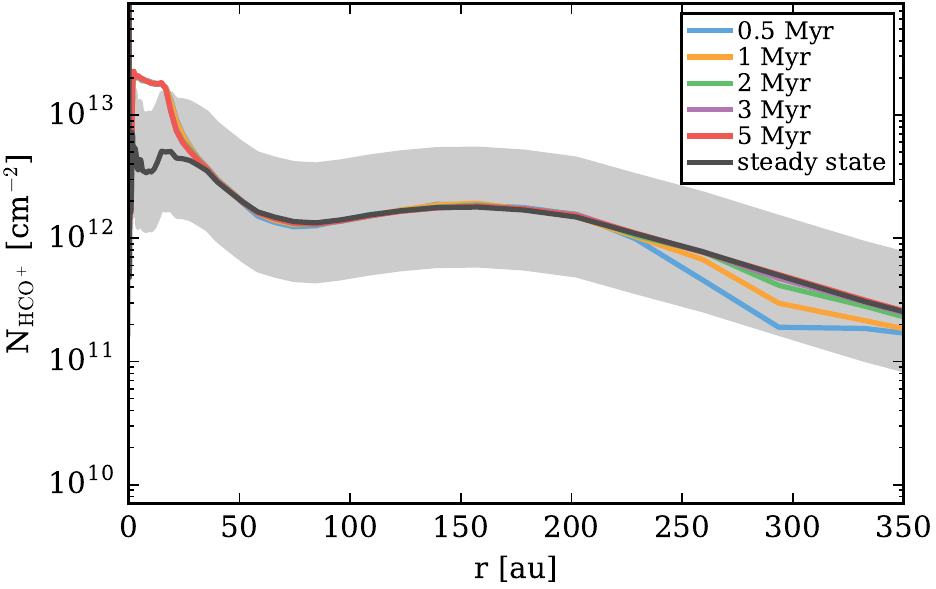}}
\resizebox{0.48\hsize}{!}{\includegraphics{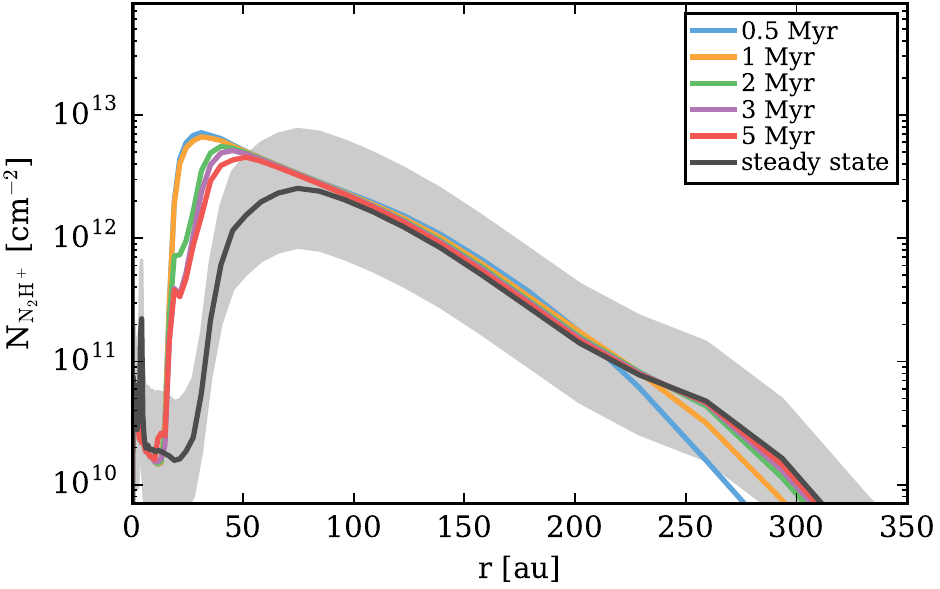}}
\caption{Comparison of time-dependent and steady-state chemistry models. Shown are the vertical column densities for HCO$^+$ and N$_2$H$^+$ at different chemical ages of the disk (colored lines) and the steady-state results (black line). The gray shaded area marks a difference of a factor 3 in the column densities relative to the steady-state model.}
\label{fig:chemtest_td}
\end{figure*}
\subsubsection{N$_2$ binding energy}
\label{sec:N2EB}
In the UMIST~2012 database $E_\mathrm{B}$(N$_2)=0.67\times E_\mathrm{B}$(CO) (see Table~\ref{table:adsenergies}). This ratio is close to the value of 0.65 used to explain observations of prestellar cores showing that N$_2$ freezes out at lower temperatures than CO \citep[e.g.][]{Bergin1997,Bergin2001b,Ceccarelli2005a}. Also disk observations of N$_2$H$^+$ indicate that $E_\mathrm{B}$(N$_2)<E_\mathrm{B}$(CO) \citep{Qi2003an,Qi2013bc}. However, the ratio of 0.65 is in contradiction with laboratory experiments which show $E_\mathrm{B}$(N$_2)\approx0.9\times E_\mathrm{B}$(CO) \citep{Oberg2005,Bisschop2006,Fayolle2016}. One explanation for this discrepancy might be differential freeze-out but it is unclear if this is sufficient \citep{Bisschop2006,Fayolle2016}. 

In our models the N$_2$H$^+$ column density drops by about one order of magnitude if we set $E_\mathrm{B}$(N$_2)=0.9\times E_\mathrm{B}$(CO) (model EBN2 in Fig.~\ref{fig:chemtest}). Contrary to our model \citet{Aikawa2015} find that in their disk model the N$_2$H$^+$ abundance structure is not strongly affected by setting $E_\mathrm{B}$(N$_2)=E_\mathrm{B}$(CO). They argue that due to the sink effect, CO is anyway depleted and therefore the N$_2$H$^+$ abundance is not strongly affected by setting $E_\mathrm{B}$(N$_2)=E_\mathrm{B}$(CO). We do not include dust surface chemistry in our model therefore the CO sink effect is less efficient. \citet{Aikawa2015} argue that the sink effect depends on various parameters (turbulence in the disk, efficiency of the conversion of CO to less volatile species) therefore they also studied a no-sink model. In this no-sink model the N$_2$H$^+$ abundance is sensitive to the binding energy of N$_2$, very similar to our models. 

The main difference between the sink and no-sink model is the location of the CO ice line and the N$_2$H$^+$ layer. In the sink model CO is also depleted at temperatures higher than its sublimation temperature, consequently the CO ice line moves to higher/warmer layers of the disk. In the no-sink model the CO ice line is given by the adsorption/desorption equilibrium for CO which depends on the chosen CO binding energy. However, with the exception of the exact location of the CO ice line and the N$_2$H$^+$ layer our models with $E_\mathrm{B}$(N$_2)=0.67\times E_\mathrm{B}$(CO) are in good agreement with the full chemical network/sink model of \citet{Aikawa2015}. We discuss the impact of the CO ice line location on our results in Section~\ref{sec:coiceline}.
\subsection{Comparison to the KIDA chemical network}
The reaction rates for nitrogen chemistry are not as well known as for  carbon/oxygen chemistry \citep[e.g.][]{Hily-Blant2010a,LeGal2014}. \citet{Wakelam2013a} reviewed a large number of important reactions for nitrogen chemistry. The new derived reactions rates are in included in the latest KIDA gas phase chemistry database release \citep{Wakelam2015}. 

We run models using the KIDA gas phase chemistry database instead of the UMIST~ 2012 database (see Table \ref{table:chemmodels}). We use the KIDA~ 2011 \citep{Wakelam2012a} and the KIDA~2014 \citep{Wakelam2015} releases. The additional chemistry included in P{\tiny RO}D{\tiny I}M{\tiny O} (e.g. X-ray chemistry) remains the same.

Fig.~\ref{fig:chemtest} shows the resulting HCO$^+$ and N$_2$H$^+$column densities for the KIDA2011 and KIDA2014 model in comparison to the reference model (UMIST 2012).  
In the KIDA2011 model the N$_2$H$^+$ column density is about a factor of three higher compared to the reference and KIDA2014 models. The reason are updated rate coefficients for the dissociative recombination reactions of N$_2$H$^+$ with electrons. The reaction rates at $\mathrm{20\,K}$ for $\mathrm{N_2H^+ + e^- \rightarrow N_2 + H}$ ($\mathrm{N_2H^+ + e^-\rightarrow NH+N}$) are a about a factor of five (three) higher in the KIDA~2014/UMIST~2012 releases than in the KIDA~2011 release. This explains the higher N$_2$H$^+$ abundance in the KIDA~2011 model. However, we find a good agreement for the KIDA~2014 and the UMIST~2012 release for both molecules HCO$^+$ and N$_2$H$^+$.
\section{Steady-state versus time-dependent chemistry}
\label{sec:timedepchem}
For our models we assume that the chemistry reaches a steady-state within typical lifetimes of disks (couple of million years). To verify this assumption we run time-dependent chemistry models for the reference models CI\_XN and CL\_XN. We find that for the species considered in this work HCO$^+$ and N$_2$H$^+$ the chemistry reaches steady-state within $\approx1\,\mathrm{Myr}$ in a large fraction of the disk. This is in agreement with the models of \citet{Aikawa2015} although they use a different, in particular larger chemical network.

In Fig.~\ref{fig:chemtest_td} we compare the vertical column densities $N\mathrm{_{HCO^+}}$ and $N\mathrm{_{N_2H^+}}$ of the steady-state model to the results of the time-dependent models at different times. For $r\gtrsim30\,\mathrm{au}$ and $t\gtrsim1\,\mathrm{Myr}$ to column densities of the steady-state and the time-dependent model are nearly identical. 
However, the steady-state model underestimates $N\mathrm{_{HCO^+}}$ and $N\mathrm{_{N_2H^+}}$ for $r\lesssim30\,\mathrm{au}$. The differences are caused by the ``sink effect'' for CO and N$_2$ (see Sect.~\ref{sec:HcopN2Hp} and \citealt{Aikawa2015}). This erosion of CO and N$_2$ and other neutral molecules in the disk midplane by the reaction with He$^+$  is a slow chemical process \citep[e.g.][]{Furuya2014g,Bergin2014,Helling2014} and is therefore ``over-estimated'' in the steady-state models. 

As N$_2$H$^+$ resides in deeper layers than HCO$^+$ the deviations in the column densities are more pronounced. As a consequence the N$_2$H$^+$ column density does not exactly trace the CO ice line in the midplane of the disk. The midplane CO ice line is located at $r\approx12\,\mathrm{au}$ in the time-dependent model, whereas in the steady-state model $N\mathrm{_{N_2H^+}}$ indicates a CO ice line at $r\approx30\,\mathrm{au}$. We note that in the low CR model CL\_XN the sink effect is less efficient due to the lower midplane ionization rate, therefore the deviations in the steady-state models for $N\mathrm{_{HCO^+}}$ and $N\mathrm{_{N_2H^+}}$ are smaller ($N\mathrm{_{N_2H^+}}$ indicates a CO ice line at $r\approx20\,\mathrm{au}$). 

Our comparison shows that for the bulk of $N\mathrm{_{HCO^+}}$ and $N\mathrm{_{N_2H^+}}$ the assumption of steady-state chemistry is well justified. In particular our conclusions concerning the impact of SP ionization are not affected by the artifacts in the steady-state models. A steady-state model requires about a factor of 10 less computational time than the time-dependent models. The use of the steady-state models allows us to study other important aspects like the impact of CO depletion (Sect.~\ref{sec:codepletion}) or the comparison of different chemical networks.
\section{The Turner model}
\label{sec:turnermodel}
In \citet{Turner2009bb} the SP ionization rate is calculated for a mean solar mass nebulae disk model. To calculate the SP ionization rate at the surface of the disk they scaled their CR ionization rate of $\mathrm{\zeta_{CR}=5\times10^{-18}\,s^{-1}}$ by a factor of $10^4(r/\mathrm{au})^{-2}$. To account for the attenuation of SPs as a function of hydrogen column density they applied the same equation as they use for Galactic cosmic rays (their Eq.~(2)). 

\begin{figure}
\centering
\resizebox{\hsize}{!}{\includegraphics{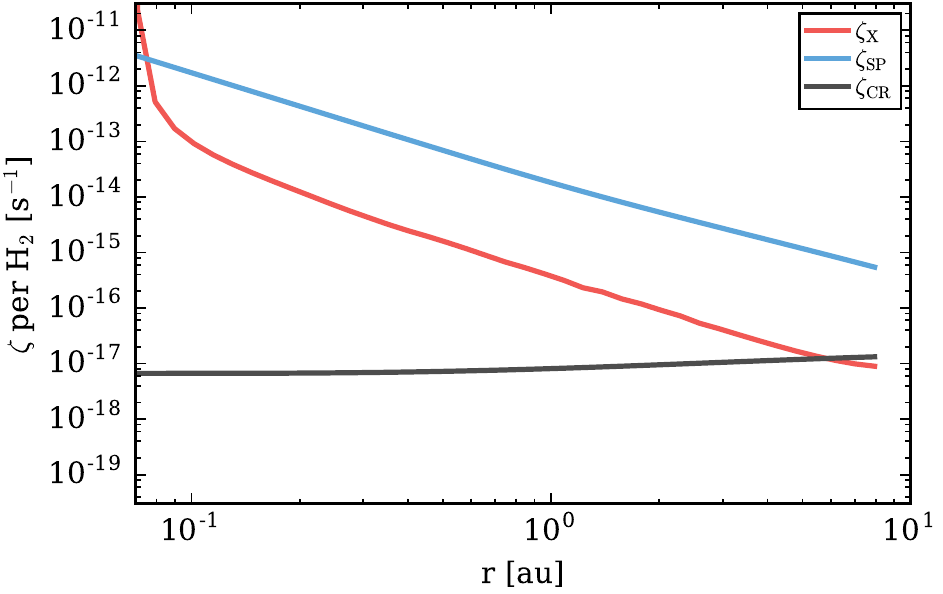}}\\
\vspace*{-0.4cm}
\resizebox{\hsize}{!}{\includegraphics{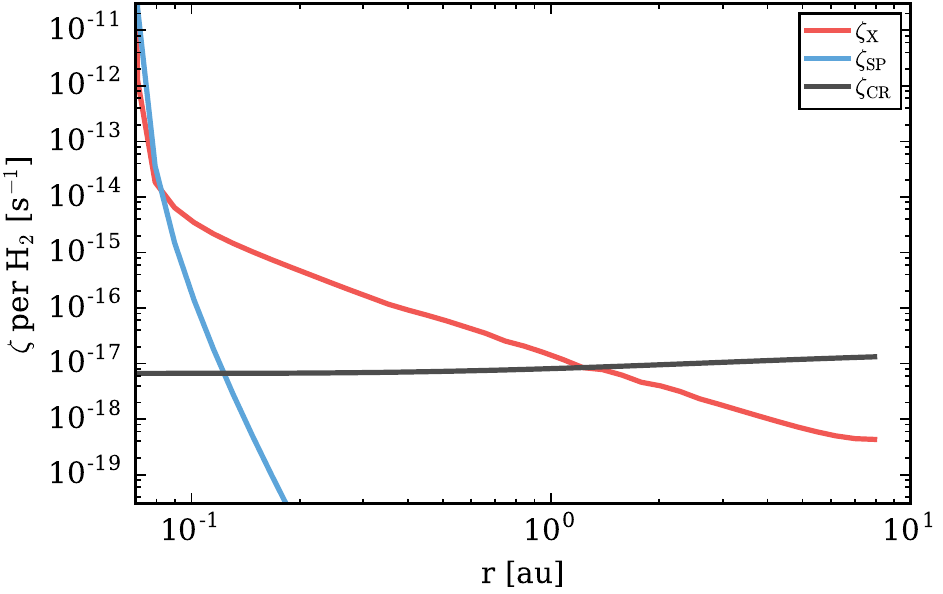}}
\caption{Ionization rates at a vertical column density of $N\mathrm{_{<H>,ver}}=3.4\times10^{24}\,\mathrm{cm^{-2}}$ ($\approx 8\,\mathrm{g\,cm^{-2}}$) as a function of distance to the star. The top panel shows the Turner model (CI\_T) the bottom panel the CI\_XN\_SP model (see Table~\ref{table:models}).}
\label{fig:ionRNH}
\end{figure}

To compare our results with the approach of \citet{Turner2009bb} we implemented their method in our model. We also used their X-ray input spectrum and applied our X-ray radiative transfer to calculate the X-ray ionization rate. In Fig.~\ref{fig:ionRNH} we show the resulting X-ray, SP and CR ionization rates as a function of radius at a vertical column density of $\approx 8\,\mathrm{g\,cm^{-2}}$ ($\approx 3\times10^{24}\,\mathrm{cm^{-2}}$) for the Turner model (CI\_T) and for our CI\_XN\_SP model. A comparison of 
Fig.~\ref{fig:ionRNH} with Fig.~1 of \citet{Turner2009bb} shows that the CI\_T model reproduces their SP ionization rates. We note that in our disk model (i.e lower disk mass) only for $r<10\,\mathrm{au}$ a vertical column density of $> 3\times10^{24}\,\mathrm{cm^{-2}}$ is reached; therefore only the inner $10\,\mathrm{au}$ are shown. 

Figure~\ref{fig:ionRNH} clearly shows the differences of the two approaches. In our models SPs can only penetrate the disk radially and therefore cannot reach the disk midplane (see Sect.~\ref{sec:diskionrates}). In the Turner model the particles can also penetrate the disk vertically down to the midplane as they use the same equation for SP attenuation as is used for Galactic CRs. We cannot reproduce their results if we allow only for radial transport of particles. The differences in the X-ray ionization rate are due to the different X-ray spectrum ($L_\mathrm{X}=2\times10^{30}\,\mathrm{erg\,s^{-1}}, T_\mathrm{X}=5.8\times10^7\,\mathrm{K}$, see Table~\ref{table:xparam}) used by \citet{Turner2009bb}.
\section{Comparison to observations}
\label{sec:cobs}
We compared the HCO$^+\,\mathrm{J=3-2}$ and N$_2$H$^+\,\mathrm{J=3-2}$ line fluxes from our reference model to the observations of \citet{Oberg2010c,Oberg2011a}. From their sample we choose the targets where both lines were detected (the six targets shown in Fig.~6 of \citealt{Oberg2011a}). The observed integrated line fluxes are in the range of $3.1-17.8\,\mathrm{Jy\,km\,s^{-1}}$ for HCO$^+\,\mathrm{J=3-2}$ and in the range of $0.4-2.9\,\mathrm{Jy\,km\,s^{-1}}$ for N$_2$H$^+\,\mathrm{J=3-2}$ (all line fluxes are scaled to a distance of 140 pc). Excluding \object{IM~Lup}, an extremely large and massive disk, from the sample gives a much narrower range of $3.1-5.4\,\mathrm{Jy\,km\,s^{-1}}$ and $0.4-1.4\,\mathrm{Jy\,km\,s^{-1}}$ for HCO$^+\,\mathrm{J=3-2}$ and N$_2$H$^+\,\mathrm{J=3-2}$, respectively.  

To calculate the line fluxes for our model we use the line transfer module of P{\tiny RO}D{\tiny I}M{\tiny O} \citep{Woitke2011} and the molecular data from the Leiden Atomic and Molecular Database \citep{Schoeier2005c,Botschwina1993,Flower1999e}. For our reference model CI\_XN we find fluxes of $2.4\,\mathrm{Jy\,km\,s^{-1}}$ and $0.78\,\mathrm{Jy\,km\,s^{-1}}$ for HCO$^+\,\mathrm{J=3-2}$ and N$_2$H$^+\,\mathrm{J=3-2}$, respectively. As the sample of \citet{Oberg2011a} is probably biased towards large and massive disks we also calculated the fluxes for a 4 times more massive disk. We find fluxes of $4.3\,\mathrm{Jy\,km\,s^{-1}}$ and $1.18\,\mathrm{Jy\,km\,s^{-1}}$ for HCO$^+\,\mathrm{J=3-2}$ and N$_2$H$^+\,\mathrm{J=3-2}$, respectively. These results are well within the range of the observations.
\end{document}